\documentclass[]{sn-jnl}% Math and Physical Sciences Reference Style
\usepackage{siunitx}
\usepackage{natbib}   %bibtex

\bibpunct{(}{)}{;}{a}{}{,}  %bibtex
\jyear{2023}%
\newcommand*\aap{A\&A}

\newcommand*\ao{Appl.~Opt.}

\newcommand*\apjl{ApJ}

\newcommand*\apjs{ApJS}

\begin{document}

\title[Science with QUVIK I]{Science with a small two-band UV-photometry mission I: Mission description and follow-up observations of stellar transients }


\author*[1]{\fnm{N.} \sur{Werner}}\email{werner@physics.muni.cz}

\author*[1]{\fnm{J.} \sur{Řípa}}\email{ripa.jakub@gmail.com}

\author[2]{\fnm{C.} \sur{Thöne}}
\author[1]{\fnm{F.} \sur{Münz}}
\author[1]{\fnm{P.} \sur{Kurfürst}}
\author[2]{\fnm{M.} \sur{Jelínek}}
\author[1]{\fnm{F.} \sur{Hroch}}
\author[3]{\fnm{J.} \sur{Benáček}}
\author[4]{\fnm{M.} \sur{Topinka}}
\author[5]{\fnm{G.} \sur{Lukes-Gerakopoulos}}
\author[1]{\fnm{M.} \sur{Zajaček}}
\author[1]{\fnm{M.} \sur{Labaj}}
\author[1,6]{\fnm{M.} \sur{Prišegen}}
\author[1]{\fnm{J.} \sur{Krtička}}
\author[7]{\fnm{J.} \sur{Merc}}
\author[8]{\fnm{A.} \sur{Pál}}
\author[9]{\fnm{O.} \sur{Pejcha}}
\author[10]{\fnm{V.} \sur{Dániel}}
\author[10]{\fnm{J.} \sur{Jon}}
\author[10]{\fnm{R.} \sur{Šošovička}}
\author[10]{\fnm{J.} \sur{Gromeš}}
\author[]{\fnm{J.} \sur{Václavík}$^{11}$}
\author[]{\fnm{L.} \sur{Steiger}$^{11}$}
\author[12]{\fnm{J.} \sur{Segiňák}}
\author[13]{\fnm{E.} \sur{Behar}}
\author[13]{\fnm{S.} \sur{Tarem}} 
\author[13]{\fnm{J.} \sur{Salh}} 
\author[13]{\fnm{O.} \sur{Reich}}
\author[14]{\fnm{S.} \sur{Ben-Ami}}
\author[15]{\fnm{M. F.} \sur{Barschke}}
\author[15,16]{\fnm{D.} \sur{Berge}}
\author[17]{\fnm{A.} \sur{Tohuvavohu}}
\author[17]{\fnm{S.} \sur{Sivanandam}}
\author[18,19,20]{\fnm{M.} \sur{Bulla}}
\author[21]{\fnm{S.} \sur{Popov}}
\author[22]{\fnm{Hsiang-Kuang} \sur{Chang}$^{22}$}

\affil[1]{\orgdiv{Department of Theoretical Physics and Astrophysics}, \orgname{Faculty of Science, Masaryk University}, \orgaddress{\street{Kotlá\v{r}ská 2}, \city{Brno}, \postcode{611 37}, \country{Czech Republic}}}

\affil[2]{\orgdiv{Astronomical Institute}, \orgname{Czech Academy of Sciences (ASU CAS)}, \orgaddress{\city{Ond\v{r}ejov}, \country{Czech Republic}}}

\affil[3]{\orgdiv{Institute for Physics and Astronomy, University of Potsdam}, \orgaddress{D-14476 \city{Potsdam}, \country{Germany}}}

\affil[4]{\orgname{INAF - Osservatorio Astronomico di Cagliari}, \orgaddress{\street{via della Scienza 5}, \city{Selargius (CA)}, \postcode{09047}, \country{Italy}}}

\affil[5]{\orgdiv{Astronomical Institute}, \orgname{Czech Academy of Sciences (ASU CAS)}, \orgaddress{\street{Bo\v{c}n\'{\i} II 1401/1a}, \postcode{141 00}, \city{Prague}, \country{Czech Republic}}}

\affil[6]{\orgdiv{Advanced Technologies Research Institute, Faculty of Materials Science and
Technology in Trnava}, \orgname{Slovak University of Technology in Bratislava}, \orgaddress{\street{Bottova
25}, \city{Trnava}, \postcode{917 24}, \country{Slovakia}}} 

\affil[7]{\orgdiv{Astronomical Institute}, \orgname{Faculty of Mathematics and Physics, Charles University}, \orgaddress{\street{V Hole\v sovi\v ck\' ach 2}, \city{Praha}, \postcode{180 00}, \country{Czech Republic}}}

\affil[8]{\orgdiv{Research Centre for Astronomy and Earth Science}, \orgname{Konkoly Observatory}, \orgaddress{\street{K Konkoly-Thege M. út 15-17}, \city{Budapest}, \postcode{H-1121}, \country{Hungary}}}

\affil[9]{\orgdiv{Institute of Theoretical Physics}, \orgname{Faculty of Mathematics and Physics, Charles University}, \orgaddress{\street{V Holešovičkách 2}, \city{Prague} \postcode{180 00}, \country{Czech Republic}}}

\affil[10]{\orgname{Czech Aerospace Research Centre (VZLU)}, \orgaddress{\city{Prague}, \postcode{199 05}, \country{Czech Republic}}}

\affil[11]{\orgdiv{Research Centre for Special Optics and Optoelectronic Systems (TOPTEC), Institute of Plasma Physics}, \orgname{Czech Academy of Sciences}, \orgaddress{\street{Za Slovankou 1782/3}, \postcode{182 00}, \city{Prague}, \country{Czech Republic}}}

\affil[12]{\orgname{PEKASAT SE}, \orgaddress{\city{Brno}, \country{Czech Republic}}}

\affil[13]{\orgname{Department of Physics, Technion}, \orgaddress{\city{Haifa 32000}, \country{Israel}}}

\affil[14]{\orgname{Department of Particle Physics and Astrophysics, Weizmann Institute of Science}, \orgaddress{Herzl St 234, \city{Rehovot}, \country{Israel}}}

\affil[15]{\orgname{Deutsches Elektronen-Synchrotron DESY}, \orgaddress{Platanenallee 6, 15738 \city{Zeuthen}, \country{Germany}}}

\affil[16]{\orgname{Institut f\"ur Physik, Humboldt-Universit\"at zu Berlin}, \orgaddress{Newtonstrasse 15, 12489 \city{Berlin}, \country{Germany}}}

\affil[17]{\orgname{Department of Astronomy \& Astrophysics, University of Toronto}, \orgaddress{50 St. George Street, \city{Toronto}, Ontario, M5S 3H4 \country{Canada}}}

\affil[18]{\orgdiv{Department of Physics and Earth Science}, \orgname{University of Ferrara}, \orgaddress{\street{via Saragat 1}, \postcode{I-44122}, \city{Ferrara}, \country{Italy}}}

\affil[19]{\orgdiv{Sezione di Ferrara}, \orgname{INFN}, \orgaddress{\street{via Saragat 1}, \postcode{I-44122}, \city{Ferrara}, \country{Italy}}}

\affil[20]{\orgdiv{Osservatorio Astronomico d’Abruzzo}, \orgname{INAF}, \orgaddress{\street{via Mentore Maggini snc}, \postcode{I-64100}, \city{Teramo}, \country{Italy}}}

\affil[21]{\orgdiv{ICTP – Abdus Salam International Center for Theoretical Physics}, \orgaddress{Strada Costiera 11, I-34151} \city{Trieste}, \country{Italy}}

\affil[22]{\orgdiv{Institute of Astronomy}, \orgname{National Tsing Hua University}, \orgaddress{\street{101 Sec. 2 Kuang-Fu Rd.}, \city{Hsinchu}, \postcode{300044}, \country{Taiwan, Republic of China}}}


\abstract{
This is the first in a collection of three papers introducing the science with an ultra-violet (UV) space telescope on an approximately 130~kg small satellite with a moderately fast re-pointing capability and a real-time alert communication system approved for a Czech national space mission. The mission, called \emph{Quick Ultra-Violet Kilonova surveyor---QUVIK}, will provide key follow-up capabilities to increase the discovery potential of gravitational wave observatories and future wide-field multi-wavelength surveys. The primary objective of the mission is the measurement of the UV brightness evolution of kilonovae, resulting from mergers of neutron stars, to distinguish between different explosion scenarios. The mission, which is designed to be complementary to the \emph{Ultraviolet Transient Astronomy Satellite---ULTRASAT}, will also provide unique follow-up capabilities for other transients both in the near- and far-UV bands. Between the observations of transients, the satellite will target other objects described in this collection of papers, which demonstrates that a small and relatively affordable dedicated UV-space telescope can be transformative for many fields of astrophysics. 
}


\maketitle

\section{Introduction}\label{sec1}

The first simultaneous detection of gravitational waves and electromagnetic radiation on 2017 August 17 
\citep{abbott2017b,abbott2017c}, resulting from a coalescence of neutron stars, marked the onset of multi-messenger astrophysics involving gravitational waves. This exciting observation showed that
neutron star mergers are of major importance for enriching the Universe with rare heavy elements such as gold and platinum.
The radioactive decay of these heavy elements powers a thermal transient at ultra-violet/visible/infrared wavelengths known as a kilonova \citep{li1998,metzger2010,metzger2019}.
To make further breakthroughs in the study of kilonovae, ultra-violet (UV) observations early after the explosion are required \citep{Arcavi2018}. In this collection of papers, we present an overview of the rich science that can be achieved with a UV space telescope on a small satellite with a moderately fast repointing capability and real-time alert communication system. We show that the \emph{Quick Ultra-Violet Kilonova surveyor (QUVIK)} two-band UV-photometry mission \citep{Werner2022}, which we describe in Sect. \ref{sec:mission}, can provide a breakthrough in our understanding of kilonovae as well as significantly expand our knowledge and discovery potential in other fields of astronomy. 

As we show in Sect. \ref{sec:kilonovae}, the early measurement of the brightness evolution of kilonovae in the UV band will allow us to distinguish between different merger scenarios. However, the ability to distinguish between different models depends critically on the capability to point to the target location and start monitoring the emission early, about an hour after receiving the gravitational wave signal. In the case of GW170817, which represents the current state of the art of kilonova observations, the kilonova AT2017gfo was discovered 11 hours after the merger and the first UV observation was performed 15 hours after the merger with \emph{Swift}/UVOT \citep{evans2017}.
The current scenarios known to produce UV/optical radiation from neutron star mergers include shock-powered, nucleosynthesis-powered, and free neutron decay-powered models,
and discriminating between them is only possible by using UV observations performed in the first hours after the merger of neutron stars \citep{metzger2019,dorsman2023}. The early UV emission, in combination with optical and near-infrared observations that can be performed from the ground, also provides a tool to probe the properties of the merger remnant, which can be a black hole, a stable massive neutron star, or a supramassive/hypermassive neutron star that after some time (which can also be inferred from the UV observations) collapses into a black hole \citep{kasen2015,metzger2019,Sarin2021}. It also allows us to determine some of the outflow parameters. This information determined from early UV photometry cannot be gleaned from observations performed later or at longer wavelengths. This provides an opportunity for truly breakthrough science, potentially allowing us to constrain the contribution of neutron star mergers to the formation of heavy, so-called r-process, elements \citep[][]{lattimer1974,wanajo2014}. 
 
The small satellite, optimised for early photometry of kilonovae, will be particularly well suited for the follow-up observations of a wide range of transient sources, such as gamma-ray bursts \citep[GRBs,][]{Piran2004,Zhang2004,Meszaros2006,Kumar2015}, supernovae \citep[SNe,][]{Bethe1990,smartt2009,2010ApJ...725..904N,rabinak2011}, outbursts in active galactic nuclei \citep{Ho2008,padovani2017}, tidal disruption of stars by supermassive black holes \citep{Dai2018,gezari2021}, etc. Gamma-ray burst science will benefit greatly from an onboard GRB detector with localisation capabilities \citep[GALI,][]{rahin2020}, enabling fast UV follow-up observations that can currently only be performed with the ageing Neil Gehrels Swift Observatory \citep[\emph{Swift},][]{gehrels2004} and soon, in the visible band, by \emph{SVOM} \citep{atteia22}. Such an onboard GRB detector will also enable the fastest possible follow-up of kilonovae that happen to coincide with short GRBs. Observations of supernovae and tidal disruption events will benefit from synergies with other major observatories, which include the Vera Rubin Observatory \citep{Ivezic2019} and the Square Kilometre Array \citep{Dewdney2009}. Importantly, follow-up observations of transients in the UV band will provide opportunities for unexpected discoveries and for the discovery of new, yet unknown, classes of transients. Between the observations of transient sources, the satellite will have an opportunity to perform observations of other targets of interest for the scientific community, such as stars and stellar systems, and galactic nuclei, described in papers II \citep{quvik2ssrv} and III \citep{quvik3ssrv} of this collection.

A particularly important synergy will be provided by the Ultraviolet Transient Astronomy Satellite \citep[\emph{ULTRASAT},][]{Ben-Ami2022,Shvartzvald2023}, which is an Israeli mission carrying a telescope with a 33\,cm aperture and a very large FoV of $200$ square degrees. It is optimised for the 230--290~nm near-ultraviolet (NUV) band and is planned to operate in a geostationary orbit. \emph{ULTRASAT} is expected to reach a 5$\sigma$ sensitivity of 22.3 AB limiting magnitude in $3 \times 300$\,s integrations. It will operate in two modes: it will stare at two regions for 6 months each, and it will perform Target of Opportunity (ToO) observations slewing on a target within 15 minutes from receiving the trigger position. Follow-up observations of \emph{ULTRASAT} targets with \emph{QUVIK} in complementary NUV and far-ultraviolet (FUV) bands with a narrower point-spread function (PSF) will help to multiply the scientific return of both missions.

\section{\emph{QUVIK} mission overview
\label{sec:mission}} 

\begin{figure}[h]
    \centering
    \resizebox{0.7\textwidth}{!}{\includegraphics{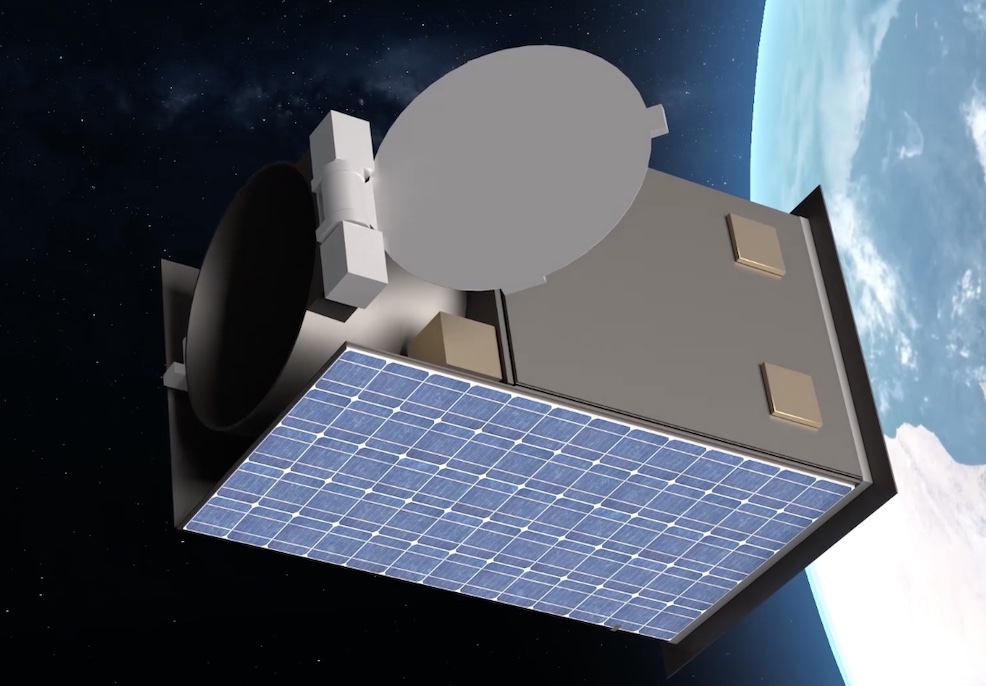}}\vspace*{2mm}
    \caption{A render of the proposed Quick Ultra-Violet Kilonova surveyor (\emph{QUVIK}).}%
    \label{fig:QUVIKimage}
\end{figure}

The primary objective of \emph{QUVIK}, illustrated in Fig. \ref{fig:QUVIKimage}, is the early measurement of the brightness evolution of kilonovae in the UV band to distinguish between different models of their explosion. This objective drives the design parameters of the mission summarised in Table \ref{table:requirements}.
Our baseline mission has a moderately large FoV of at least $\SI{1.0}{\degree} \times \SI{1.0}{\degree}$, a PSF of $\lesssim \SI{2.5}{arcsec}$ (FWHM), and a NUV photometric sensitivity of $22$ AB magnitude in less than 1000 second at signal-to-noise ratio (SNR) of $5$, and observation start latency for an unocculted target (ToO response time) better than $\SI{20}{minutes}$. Its NUV spectral range of $\sim \SI{260}{\nano\metre}$ to $\sim \SI{360}{\nano\metre}$ is the primary band for kilonova observations and provides complementarity to \emph{ULTRASAT} at longer wavelengths. The FUV band spanning from $\sim \SI{140}{\nano\metre}$ to $\sim \SI{190}{\nano\metre}$ will give \emph{QUVIK} unique capabilities as this band is currently only covered by UVIT on \emph{AstroSat} \citep{singh2014}.\footnote{On August 18 2022, NASA announced the selection of the Ultraviolet Explorer \citep[UVEX,][]{Kulkarni2021} as one of two MIDEX proposals for a detailed study. A final down-selection will happen in 2024 with the launch expected sometime after 2028, likely around 2030.  UVEX, which is a significantly larger mission than \emph{QUVIK}, would cover the same spectral bands and, if selected, the mission would most likely launch after our proposed small satellite. \emph{QUVIK} could thus serve as a pathfinder for UVEX and even if operated simultaneously, it would provide a large discovery potential in the very rich field of UV astronomy. } Close coordination of the FUV-capable \emph{QUVIK} with \emph{ULTRASAT}, will thus multiply the scientific returns of both missions.

\begin{table}[h]
\caption{Baseline \emph{QUVIK} mission parameters.}%
\label{table:requirements}
\centering
\begin{tabular}{lc}
  \hline \hline
  \emph{QUVIK} & Requirement \\
  \hline
  \hline
  NUV FoV                                  & $\geq \SI{1.0}{\degree} \times \SI{1.0}{\degree}$ \\
  Angular resolution (FWHM)                          & $\lesssim 2.5$~arcsec\\
  NUV bandpass                                       & $\sim 260$--$\SI{360}{\nano\metre}$ \\
   FUV bandpass                                       & $\sim 140$--$\SI{190}{\nano\metre}$ \\
  NUV photometric sensitivity                        & 22 AB mag (SNR 5 in 1000\,s)$^a$ \\
  FUV photometric sensitivity                       & 20 AB mag (SNR 5 in 1000\,s) \\
  Observation start latency (for unocculted targets) & $< \SI{20}{\minute}$ \\
  Baseline mission duration                          & $3$~years \\
  Target launch date                                 & 2028--2029 \\
  \hline
\end{tabular}
{\noindent \tiny $^a$For a kilonova on the background of its host galaxy at an offset of $1.5$ effective radii.}
\end{table}

\subsection{Orbit}
The most cost-effective orbit, with many launch opportunities for small-satellite missions, is a Sun-synchronous (SSO) low-Earth orbit. To minimise thermal cycling, the preferred orbit for \emph{QUVIK} is the dawn-dusk SSO (with a mean local time at the ascending node of approximately 6:00 AM/PM) where the spacecraft is orbiting the Earth close to its terminator and the Sun appears approximately normal to the orbital plane. 
The currently baselined orbital altitude of $\SI{550}{\kilo\metre}$ implies that the spacecraft is illuminated permanently for about $9$ months, whereas in the remaining time of the year it is eclipsed for up to $22$ minutes per orbit. 

\subsection{Spacecraft}

\begin{figure}[h]
    \centering
    \includegraphics[width=7cm]{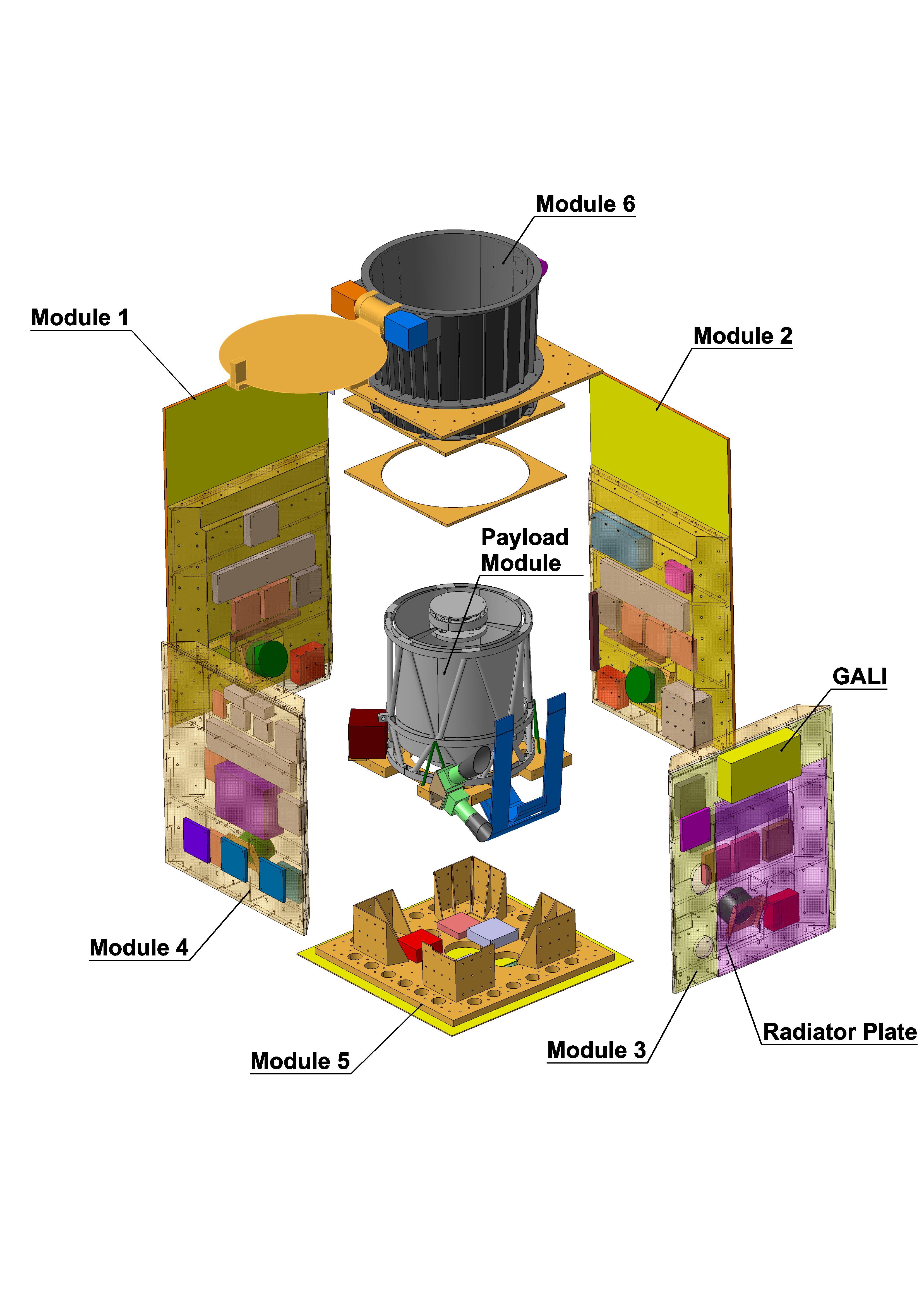}
    \caption{The modular design of \emph{QUVIK} with the telescope placed in the centre of the spacecraft and the platform distributed around it in six individual blocks. }
    \label{fig:QUVIK_spacecraft}
\end{figure}

The design of the spacecraft is based on a modular small-satellite platform with the primary telescope payload in its centre and the platform distributed along the side walls. The configuration indicated in Fig.~\ref{fig:QUVIK_spacecraft}, shows the platform distributed in four modules/blocks (Modules 1--4), the spacecraft adapter/separation system at the bottom (Module 5), and the module with the baffle and telescope front door at the top (Module 6). The telescope is mounted via three bipods to minimise the transfer of mechanical stresses and vibrations from the spacecraft or launcher.

The spacecraft has 3 solar panels (Modules 1, 2 and around the spacecraft adapter/separation system on Module 5), which can generate between $\SI{64}{\watt}$ and $\SI{200}{\watt}$, depending on the orientation with respect to the Sun. The total power consumption of all subsystems is up to $\SI{100}{\watt}$. The observation planning will be driven by the total power, which needs to be generated to maintain the spacecraft subsystems during the observations.

The spacecraft is stabilised in 3 axes by 4 reaction wheels in a pyramidal configuration. The reaction wheels are continuously offloaded by 3
perpendicular magnetorquers. Attitude knowledge is provided by the gyro-stellar estimator fusing information from two onboard star trackers and a navigation-grade fibre-optic gyroscope. A magnetometer is planned to be used to support commanding magnetic
torquers and as a safe-mode attitude sensor. Eight coarse sun-sensors are used in safe mode, guaranteeing high robustness, low power consumption, and the ability to determine the Sun vector in
almost any spacecraft condition. A Global Navigation Satellite System (GNSS) receiver is intended to be used for onboard time synchronisation and as a position sensor.

To achieve our goal for the pointing stability of $\SI{2.5}{arcsec}$ within a $\SI{60}{\second}$ exposure and sufficient agility for fast re-pointing (our goal of $< \SI{20}{\minute}$ observation start latency requires at least $\SI{0.4}{\degree \per \second}$), the spacecraft is designed as compact and rigid as possible, without antenna booms or deployable solar panels. 

The majority of science data processing will be performed on the ground, but the satellite is also expected to perform several processing tasks onboard. The tasks performed onboard include combining shorter exposures of the same field into one image with a higher SNR; computing metrics and meta-data for image evaluation, including image orientation, level of cosmic-ray contamination, and possible blurring of PSF due to Attitude and Orbit Control System (AOCS) instability; creating cut-outs from the original full-frame images. These tasks will significantly reduce the amount of data for downlink to ground and improve the image quality.

The satellite will use an S-band radio for telemetry, tracking, and command uplink and downlink. For redundancy, two S-band radios and two patch antennas will be used on opposite sides of the spacecraft. An X-band radio will be used for data downlink, allowing to transfer 400 full images ($\SI{32}{\mega\byte}$ each) twice a day. A dedicated L-band radio will be implemented for near real-time bi-directional communication through a geostationary satellite constellation.

The satellite will by default follow an observing schedule of non-transient targets predetermined by the science operation centre (SOC). Once a gravitational wave or electromagnetic transient, which fulfils the criteria for target-of-opportunity (TOO) observations, is identified by the SOC, a command will be sent to the satellite to terminate the ongoing observation and slew on the TOO target. The requirement is that the satellite receives the alert within 5 minutes after it has been issued by the SOC and will slew on target and commence observation within 15 minutes after the alert reception. While the instantaneously accessible sky fraction on low-Earth orbit is 43\%, during the 90 minute orbit the satellite can access nearly 80\% of the sky (see Table \ref{table:telescope}).

GW observatories are expected to provide alerts with rough parameter estimates within minutes and refined parameters within hours. \emph{ULTRASAT}, with its fast slewing capability, a considerable fraction of instantaneously accessible sky, and very large FoV, is expected to quickly detect and localise a substantial fraction of KNe out to a distance of at least 300$\, \text{Mpc}$ and within $15 \, \text{minutes}$ distribute an alert to the astronomical community \citep{Shvartzvald2023}. Follow-up observations with \emph{QUVIK} will provide two-band photometry in complementary spectral bands and with a higher spatial resolution. The PSF of 2.5$\, \text{arcsec}$ will enable us to better isolate KNe on the background of their host galaxies.
Following the GW alert with the initial localisation, and before a more precise localisation is established, \emph{QUVIK} will start mosaic observations of the potential target area. The mosaic observations will be performed in coordination with other observatories/teams to minimise unnecessary search overlap and find the counterpart as quickly and efficiently as possible, e.g. employing the Gravitational Wave Treasure Map tool\footnote{\url{http://treasuremap.space}} \citep{Wyatt2020}.

\subsection{Science payloads}

\subsubsection{The UV-telescope and detectors
\label{sect.payload}}

\begin{figure}[h]
    \centering
    \includegraphics[width=5.5cm]{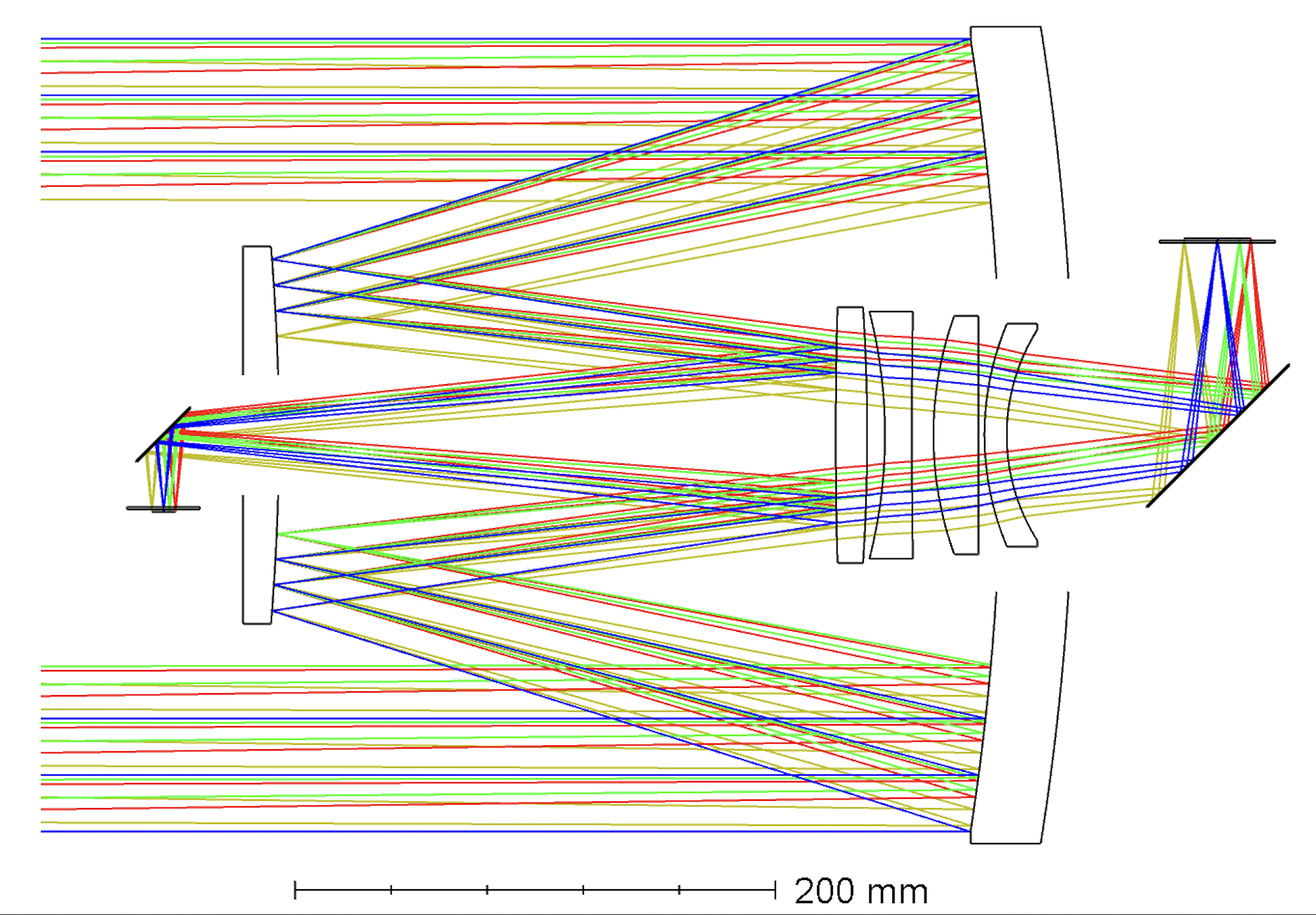}
      \includegraphics[width=5.5cm]{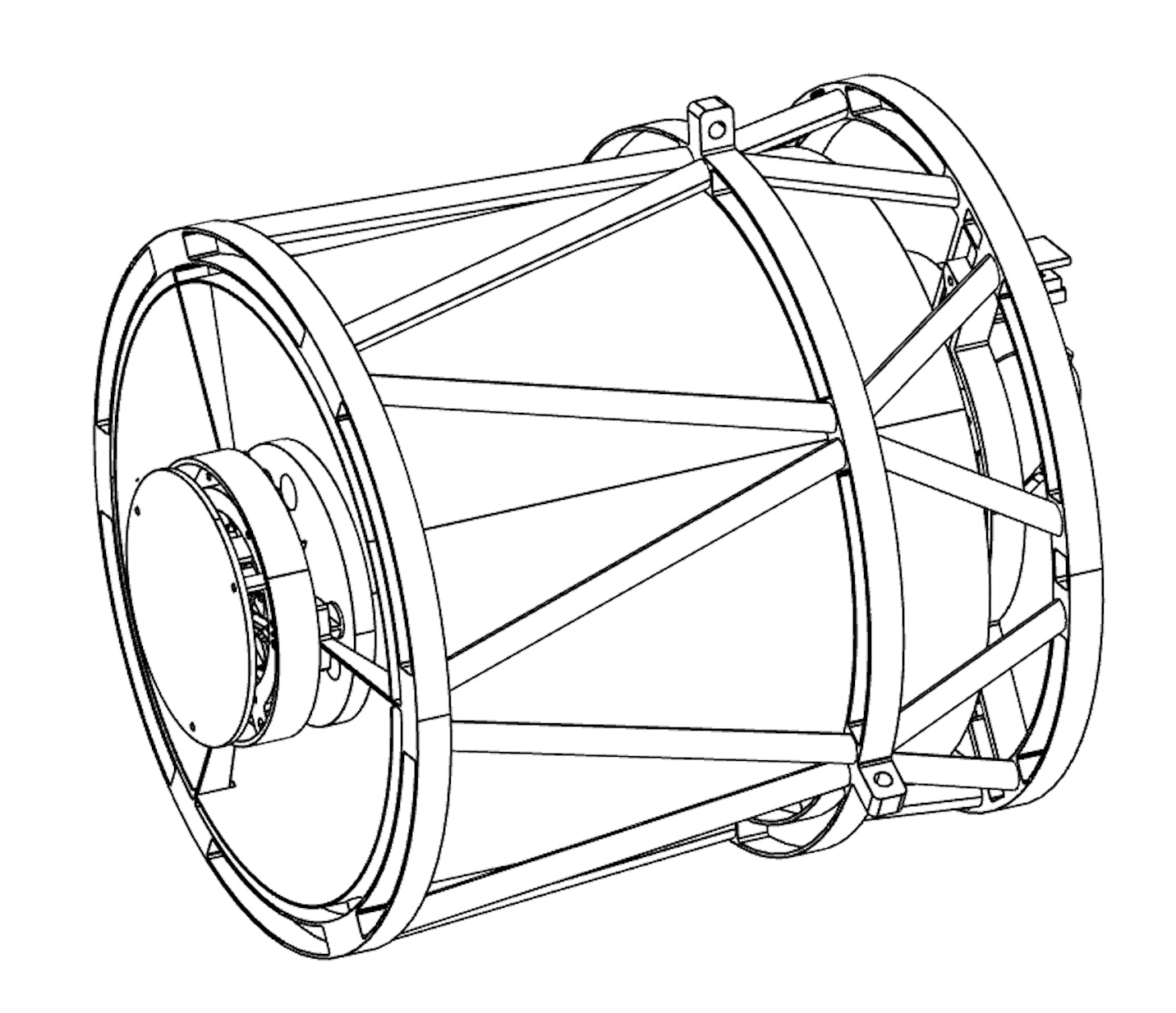}
    \caption{The optical design of the \emph{QUVIK} telescope payload. Colours represent light rays from different parts of the FoV. The NUV focal plane is behind the primary mirror. The corrector contains 4 lenses made of fused silica, magnesium fluoride and calcium fluoride. The corrector has an aspheric shape and the first surface of the first lens also serves as a beam splitter. The FUV optical path is formed by the primary and secondary mirrors with the front surface of the first lens of the corrector. The FUV focal plane is behind the secondary mirror. }
    \label{fig:optical design}
\end{figure}

The primary payload consists of a modified Cassegrain telescope with a 33~cm diameter primary mirror (Fig. \ref{fig:optical design}). A dichroic mirror splits the light into NUV and FUV channels that are simultaneously imaged by two focal planes.
The NUV channel path is a classic 2-mirror telescope with a field corrector and a FoV of $1^\circ\times1^\circ$ imaged by a 4k$\times$4k complementary metal–oxide–semiconductor (CMOS) sensor. The sensors being considered for \emph{QUVIK} are the GSENSE4040BSI\footnote{\url{https://www.gpixel.com/en/pro_details_11100.html}} and the UV-detector developed for \emph{ULTRASAT} \citep{asif2021,BastianQuerner2021,liran2022}. 
The corrector contains 4 lenses of fused silica, magnesium fluoride and
calcium fluoride. It has an aspheric shape and the first surface of the
first lens also serves as a beam splitter. The FUV optical path is formed by the primary and
secondary mirrors with the front surface of the first lens of the corrector. The FUV focal
plane is behind the secondary mirror. The FoV of the FUV channel is $\SI{0.25}{\degree} \times \SI{0.25}{\degree}$ and is imaged by a CMOS sensor. The quantum efficiency of modern backside illuminated (BSI) CMOS detectors enables large improvements with respect to previously flown UV missions  \citep[e.g.][]{nikazd2017}. After accounting for losses in the optical system, taking into account the properties of mirror and corrector coatings, as well as focal plane bandpass filters and the quantum efficiency of the preselected CMOS sensor, the final effective area in the NUV band will be $140 \textup{--} \SI{180}{\square\centi\metre}$. The expected effective area in the FUV band, after accounting for all the losses, will be $20 \textup{--} \SI{25}{\square\centi\metre}$ (see Table \ref{table:telescope} for a summary of telescope parameters).

\begin{table}[h]
\caption{Baseline telescope parameters.}%
\label{table:telescope}
\centering
\begin{tabular}{lc}
  \hline
  \hline
  Physical collection area                                      & 635 cm$^{2}$ \\
  Effective area                               & NUV:~$140 \textup{--} \SI{180}{\square\centi\metre}$ \\
                                     & FUV:~$20 \textup{--} \SI{25}{\square\centi\metre}$ \\
   NUV channel FoV            & $\qtyproduct{1 x 1}{\degree}$ %($1.41^\circ$) 
  \\
  NUV channel focal length                        & $\SI{1064}{\milli\metre}$ \\
  FUV channel FoV & $\qtyproduct{0.25 x 0.25}{\degree}$ 
  \\
  FUV channel focal length                                & $\SI{1212}{\milli\metre}$ \\
  Optical diameter                                 & $\SI{330}{\milli\metre}$ \\
   Sun avoidance angle                       & 50$^\circ$ \\
  Earth limb avoidance angle                & 20$^\circ$ \\
  Moon avoidance angle                      & 20$^\circ$ \\
  Instantaneously accessible sky fraction   & 0.43 \\
  Per-orbit accessible sky fraction         & 0.79 \\
  \hline
\end{tabular}
\end{table}

The detectors require cooling to a temperature of $\SI{-30}{\celsius}$ in order to reach a dark current of 
$0.1$\,e$^-$s$^{-1}$px$^{-1}$, which is needed to reduce the noise.
To maintain the required photometric accuracy, the detectors need to be kept at a stable, selected temperature for which the pixel gains were calibrated.
The preferred cooling method is passive heat removal by a heat-strap connected to the spacecraft radiator. The detector temperature will be stabilised using a resistive heater.

The sensitivity, i.e. the SNR of an observed point source, has been calculated as a function of the total accumulation time for co-added images with shorter individual exposures in the NUV band of 260--360\,nm following the standard charge-coupled device (CCD) equation \citep{howell2000,keller2015}:

\begin{equation}
SNR=\frac{n_{*}t}{\sqrt{n_{*}t+n_\mathrm{pix}\left( n_\mathrm{ZL}t+n_\mathrm{G}t+n_\mathrm{D}t+n_\mathrm{S}t+N_\mathrm{ex}N_\mathrm{R}^{2}\right)}},
\end{equation}

\noindent where the total accumulated exposure $t=N_\mathrm{ex} t_\mathrm{ex}$ is obtained by co-adding $N_\mathrm{ex}$ number of short exposures $t_\mathrm{ex}$. We have assumed the effective area of the telescope, including the quantum efficiency of the GSENSE4040BSI sensor (one of the candidates for the \emph{QUVIK} detector) and the transparency of the optical path, to be $\SI{180}{\square\centi\metre}$. The other parameters are:

\begin{itemize}
    \item $n_\mathrm{*}$ = 1.27\,e$^-$\,s$^{-1}$ for 21 AB mag and 0.504\,e$^-$\,s$^{-1}$ for 22 AB mag are the detected photo-electron rates from the source in the band of $260 \textup{--} \SI{360}{\nano\metre}$ (accounting for the optical throughput of the telescope and the quantum efficiency of the NUV sensor).
    \item $n_\mathrm{pix}$ = 4 ($2\times2$ pixels) is the number of pixels covering the PSF assuming a pixel resolution of 1.75\,arcsec\,px$^{-1}$.
    \item $n_\mathrm{ZL}$ = 0.03\,e$^-$s$^{-1}$px$^{-1}$ is the detected zodiacal light background as the photo-electron rate for the $260 \textup{--} \SI{360}{\nano\metre}$ band and for a typical brightness of 100 S$_{10}$(V) \citep{Leinert1975,Levasseur-Regourd1980}.
    \item $n_\mathrm{G}$ = 0.06\,e$^-$s$^{-1}$px$^{-1}$ is the detected host galaxy background as the photo-electron rate for a surface brightness of 25\,mag\,arcsec$^{-2}$ for the offset of 1.5\,$R_\mathrm{e}$ ($R_\mathrm{e}$ is the effective galaxy radius) of a kilonova from the centre of the galaxy, which is the median offset of short GRBs from their host galaxies.
    \item $n_\mathrm{D}$ = 0.1\,e$^-$s$^{-1}$px$^{-1}$ is the dark current of the GSENSE4040BSI sensor at $\SI{-30}{\celsius}$.
    \item $n_\mathrm{S}$ = 1.0\,e$^-$s$^{-1}$px$^{-1}$ is the photo-electron background rate due to stray-light.
    \item $N_\mathrm{R}$ = 3.9\,e$^-$ is the readout noise of the GSENSE4040BSI sensor.
\end{itemize}

\noindent For these parameters, the effective area in the NUV band is sufficient to detect a 22 AB magnitude object with a SNR=5 using a 900\,s (15\,min) exposure image stacked from individual $t_\mathrm{ex}=20$\,s exposures. In the case of good stability and low jitter (according to the current estimates, the jitter will be better than 1.8\,arcsec over $\SI{60}{\second}$), the individual exposures will be extended to at least $t_\mathrm{ex}=60$\,s, reducing the time to reach magnitude 22 AB to about 720\,s (12\,min). The AB magnitude of 23 will be achievable in 5460\,s (91\,min) by stacking individual 20\,s exposures and in 4200\,s (70\,min) by using $t_\mathrm{ex}=60$\,s sub-exposures. Because the satellite will be placed in low-Earth orbit, where most objects are only continuously observable for a few tens of minutes, reaching the AB magnitude of 23 will require at least two orbits of \emph{QUVIK} (one orbit takes $\sim \SI{90}{minutes}$). 

To estimate the FUV sensitivity, we have assumed the effective area of the telescope, including the estimate of the quantum efficiency of the GSENSE2020BSI sensor\footnote{\url{https://www.gpixel.com/en/pro_details_1194.html}} \citep{gill2022}, a candidate for the FUV detector, and the transparency of the optical path, to be $\SI{20}{\square\centi\metre}$. The other parameters are:

\begin{itemize}
    \item $n_\mathrm{*}$ = 0.35\,e$^-$\,s$^{-1}$ for 20 AB mag is the detected photo-electron rate from the source in the $140 \textup{--} \SI{190} {\nano\metre}$ band (accounting for the effective area).
    \item $n_\mathrm{pix}$ = 9 ($3\times3$ pixels) is the number of pixels covering the PSF assuming a pixel resolution of 1.11\,arcsec\,px$^{-1}$.
    \item $n_\mathrm{D}$ = 0.1\,e$^-$s$^{-1}$px$^{-1}$ is the assumed dark current of the GSENSE2020BSI sensor at $\SI{-30}{\celsius}$.
    \item $n_\mathrm{S}$ = 0.05\,e$^-$s$^{-1}$px$^{-1}$ is the assumed photo-electron background rate due to stray-light.
    \item $N_\mathrm{R}$ = 2.67\,e$^-$ is the readout noise of the GSENSE2020BSI sensor \citep{gill2022}.
    \item The zodiacal light background $n_\mathrm{ZL}$ as well as the host galaxy background $n_\mathrm{G}$ are negligible in the FUV band.
\end{itemize}

\noindent For these parameters, the effective area in the FUV band is sufficient to detect a 20 AB magnitude object with a SNR=5 using a 1000\,s exposure image stacked from individual $t_\mathrm{ex}=20$\,s exposures.

Given the limits for the NUV stray-light contribution to the background, the pointing direction for observations performed at nominal sensitivity will have to be $> \SI{50}{\degree}$ from the Sun, $> \SI{20}{\degree}$ from the Earth limb, and $> \SI{20}{\degree}$ from the Moon (see Table \ref{table:telescope}). 

\begin{figure}[h]
    \centering
    \includegraphics[width=7cm]{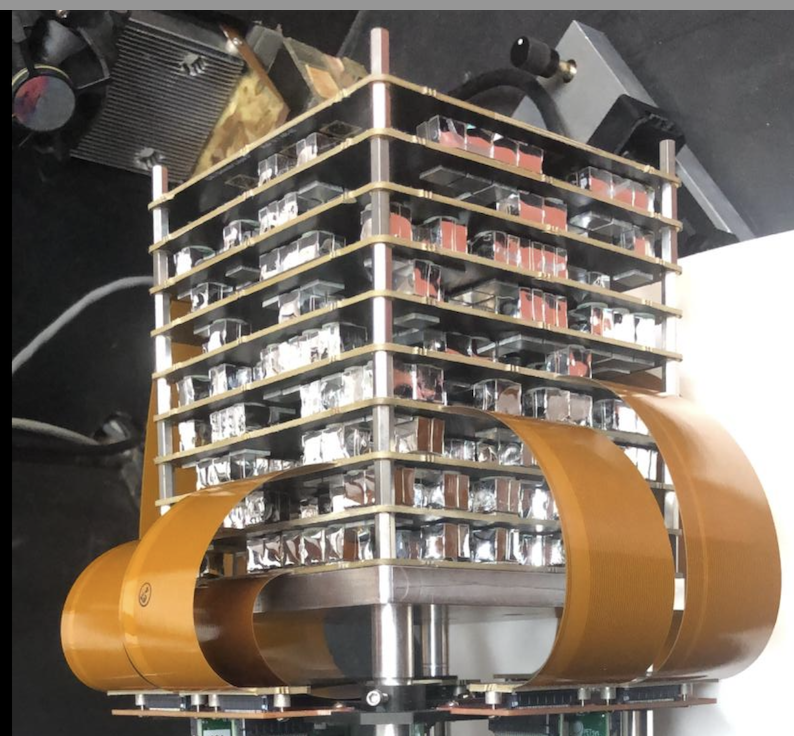}
    \caption{Engineering model of GALI proposed as the GRB detector for \emph{QUVIK}. This detector includes 362 CsI(Tl) scintillator crystals. The individually wrapped scintillators are scattered in a 3-D configuration to exploit their mutual shadowing for burst localization. }
    \label{fig:GALI_EM}
\end{figure}

\subsubsection{The gamma-ray burst detector}
\label{sect:GALI}

\emph{QUVIK} is expected to host a secondary payload, a gamma-ray burst (GRB) detector capable of localising an average GRB with a few degrees accuracy. The detector concept is the Gamma-ray burst Localizing Instrument \citep[GALI,][]{rahin2020}, which is developed by the Technion (Israel). GALI uses a novel approach of a 3-D active coded mask collecting signals from hundreds of small scintillators.
Localisation is achieved from the mutual shadowing between scintillators.
Since most scintillators are hidden inside the detector assembly, they benefit from a low background and produce a source signal only for GRBs in specific directions.
The compact configuration is made possible by Si-PM technology.
Since the scintillators serve as both detecting units and a mask, GALI provides relatively uniform coverage and sensitivity of the entire observable sky.

The GALI method is scalable and can be built to practically any size. 
A larger detector will naturally be more sensitive and also enable more accurate localisation. 
We are simulating several configurations and locations on the \emph{QUVIK} satellite, seeking the best performance for the mission, while also considering engineering constraints. An engineering model with 362 $\SI{1}{\cubic\centi\metre}$ scintillators, which is currently being tested in the laboratory, is shown in Fig.\,\ref{fig:GALI_EM}.
The GRB localisation accuracy will eventually depend on the burst flux and the detector size, with a goal of $\sim \SI{1}{\degree}$ for bright bursts and better than $\SI{10}{\degree}$ for the faintest ones still detected.
 The GRB detector will provide an onboard trigger and accurate localisation information, allowing \emph{QUVIK} to autonomously slew to the GRB location and start a measurement sequence with the UV telescope. The observing sequence could include tiled observations for GRBs where the localisation accuracy is not sufficient for a single pointing. The exact set of conditions for triggering a GRB follow-up is not yet determined and will be clarified in the following stages of the mission development.

The potential angular resolution of GALI was simulated with the MEGAlib tool \citep{zoglauer2006}, comparing two detector configurations; one with 9 layers and 362 scintillators, and another with a $2\times5$-layer configuration and 406 scintillators. The first configuration corresponds to the $12.5 \times 12.5 \times 12.5 \, \text{cm}^3$ prototype shown in Fig. \ref{fig:GALI_EM} and the second one, occupying a volume of $25\times12.5\times7 \, \text{cm}^3$, to the model in the current design of the satellite shown in Fig. \ref{fig:QUVIK_spacecraft}. We assume that the effective area of the detector is zero for the half of the sky blocked by the satellite. Since part of the field of view is also blocked by the Earth, the visible area of the sky is reduced to $\sim 1.6 \pi$ sr. For the rest of the sky, the effective area of the 9-layer configuration ranges from 56-107 cm$^2$ with an average of 90 cm$^2$. The $2\times5$-layer detector is less uniform, but overall more sensitive, with an effective area of 39-152 cm$^2$ and an average of 100 cm$^2$. 
We simulated 120\, 000 bursts in random positions using generic GRB spectra assuming a Band function with a constant photon flux of 5 ph\, s$^{-1}$\, cm$^{-2}$ in the 10--1000 keV band \citep[see][]{rahin2020}. The mean energy in this band is 87 keV, which results in a conversion from photon flux to energy flux of 1 ph\,s$^{-1}$\, cm$^{-2}$ to $1.39\times10^{-7}$ erg\, s$^{-1}$\, cm$^{-2}$.
The bursts were simulated
along with the expected sky background
considering burst durations of 1, 5, 10, 20, and 40 s. 
The GRB direction was then estimated by comparing the distribution of detected photons on the scintillators to a calibration-based sky map prepared in advance.
The accuracy statistics are presented in Table \ref{table:GALI}. We estimate that GALI will detect around 80 GRBs per year of which $\sim$12 will be short ones. For around 10 GRBs per year, we expect to achieve a better than 1-degree localisation.

\begin{table}[h]
\caption{Estimated directional errors for GRBs with a $10 \textup{--} 1000$ keV flux of $5 \, \text{ph} \, \text{s}^{-1} \, \text{cm}^{-2}$, averaged over the sky based on 120\,000 simulated bursts in random positions. GALI (top rows) refers to the current design of 362 scintillator cubes occupying a $12.5 \times 12.5 \times 12.5 \, \text{cm}^3$ volume. The $2\times5$-layer configuration refers to an alternative design of two sets of 5 layers with 406 scintillator cubes, occupying a $25\times12.5\times7 \, \text{cm}^3$ volume. The mean error refers to the mean of the deviations of the reconstructed positions from the true locations. 
 }%
\label{table:GALI}
\centering
\begin{tabular}{cccccc}
  \hline
  \multicolumn{6}{c}{GALI with 362 detectors}\\
    \hline
Time [$\unit{\second}$] & 1 & 5 & 10  & 20 & 40 \\
Detected photons   & 444.9  & 2224 & 4449  & 8898 & 17796  \\
Background photons & 2065 & 10324 & 20647 & 41295 & 82590 \\
Mean error [$\unit{\degree}$] & 4.781 & 1.93 & 1.346 & 0.963 & 0.723 \\
90\% Range [$\unit{\degree}$] & $0.97 \textup{–} 11.48$ & $0.42 \textup{–} 4.20$ & $0.32 \textup{–} 2.86$ & $0.32 \textup{–} 2.86$ & $0.16 \textup{–} 1.48$ \\
\hline
  \multicolumn{6}{c}{$2 \times 5$ layers with 406 detectors}\\
\hline
Detected photons  & 613 & 3067 & 6134 & 12268 & 24536 \\
Background photons & 2808 & 14042 & 28084 & 56169 & 112338 \\
Mean Error [$\unit{\degree}$] & 4.527 & 1.785 & 1.272 & 0.957 & 0.723 \\
90\% Range [$\unit{\degree}$] & $0.85 \textup{–} 11.39$ & $0.4 \textup{–} 3.87$ & $0.28 \textup{–} 2.72$ & $0.22 \textup{–} 2.72$ & $0.16 \textup{–} 1.51$ \\
\hline
\end{tabular}
\end{table}

\section{Kilonovae}%
\label{sec:kilonovae}

Mergers of binary neutron stars (BNS) result in a large amount of ejected neutron-rich material where heavy elements with $Z>26$ are synthesised via rapid neutron capture nucleosynthesis, the so-called r-process \citep{lattimer1974,wanajo2014}. Radioactive decay of this freshly produced material results in a transient source called kilonova \citep[KN,][]{li1998,kulkarni2005,metzger2010,metzger2019}. The current state-of-the-art in their observational studies was provided by the detection of the kilonova AT2017gfo, which resulted from the BNS merger that produced the gravitational wave source GW170817 \citep{abbott2020a}. The following large observing campaign revealed that the merger indeed resulted in a KN, the emission of which initially peaked in NUV and over about 10~days evolved redward.

\begin{figure}[h]
    \centering
    \resizebox{0.8\textwidth}{!}
    {\includegraphics{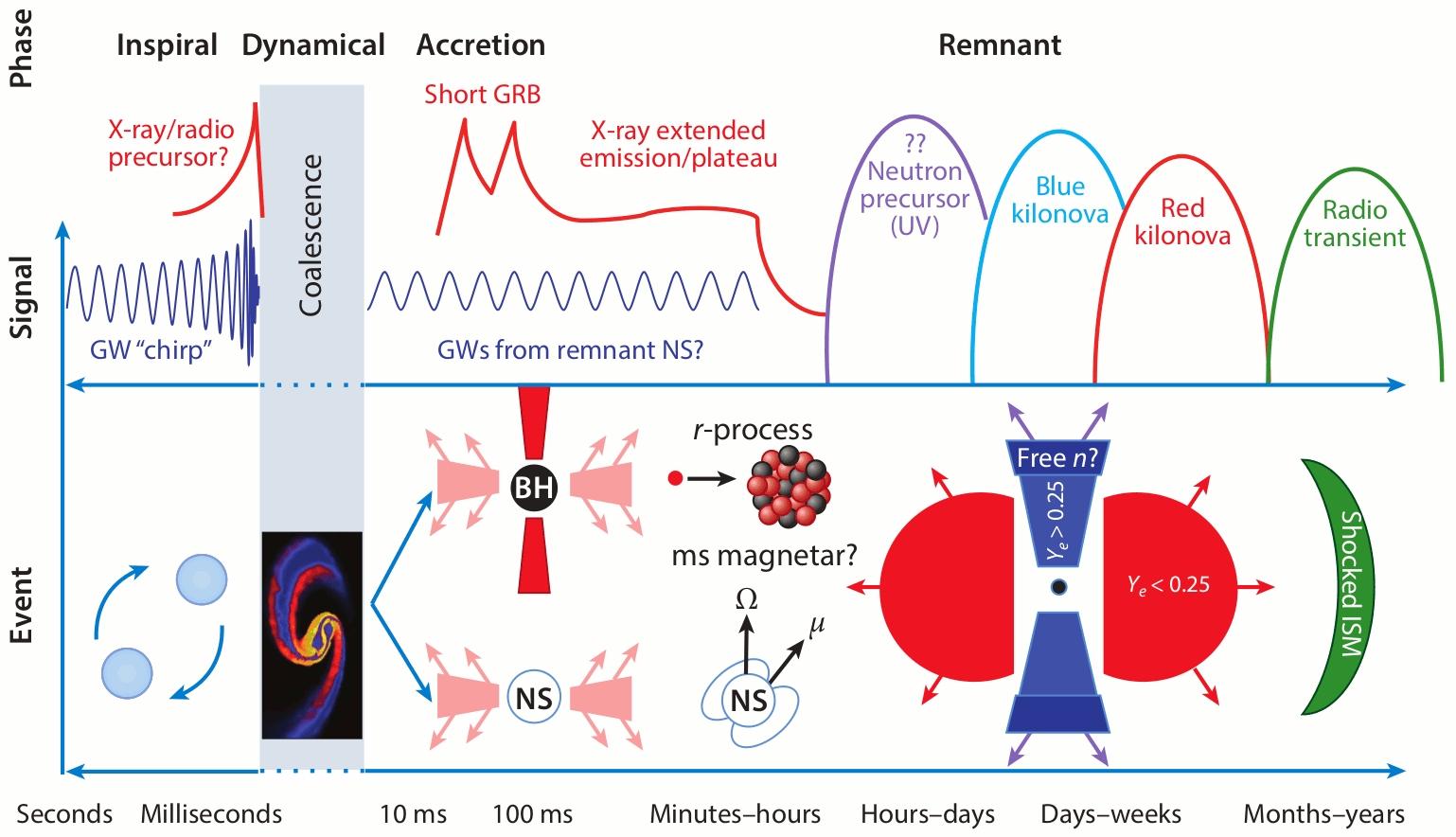}}\vspace*{2mm}
    \caption{The top shows the phases of an BNS merger as a function of time, showing the associated observational GW and EM signatures.  The bottom shows the physical phenomena producing the above signatures \cite[adopted from][]{fernandez2016}. }
   \label{fig:merger_phases}
\end{figure}

Fig.~\ref{fig:merger_phases} from \citet{fernandez2016} shows the phases of a BNS merger and its observational signatures. The final stages of the inspiral result in a gravitational wave signal and may also produce an electromagnetic (EM) precursor. The coalescence leads to the ejection of $10^{-4}$--$10^{-2}$~$M_\odot$ of unbound material \citep{metzger2012}. The ejected matter that remains gravitationally bound to the resulting compact object, which may be a massive neutron star or a black hole, falls back and forms an accretion disc that launches relativistic jets, which produce a GRB. While the GRB can only be observed when viewed close to the jet axis, the emission of the KN is nearly isotropic, with the spectral properties possibly changing depending on the viewing angle \citep[e.g.][]{Kawaguchi2018, Darbha2020, Korobkin2021, Collins2023, bulla2023a}. The ejecta is expected to be rich in newly created heavier nuclei, the decay of which results in emission that turns quickly from UV, through blue and red, to infrared. As discussed in the next section, the source of the early UV and blue emission in the kilonova AT2017gfo is not yet clear \citep{Arcavi2018}.

Gravitational wave (GW) observatories have recently also discovered another BNS merger candidate GW190425 as well as three neutron star---black hole (NS-BH) coalescences \citep{abbott2023}, but follow-up observations did not reveal any electromagnetic counterparts for these events \citep{anand2021}. Based on theoretical arguments, we would expect that, for sufficiently small mass ratios, NS-BH mergers will also produce bright KNe \citep{li1998,kawaguchi2016}. Thus, follow-up observations of these exciting events present important discovery potential. 

While GW170817 was quickly followed by a short GRB seen at an angle of $\SI{19}{\degree} \textup{--} \SI{25}{\degree}$ from the jet axis \citep{Mooley2022}, it is likely that most BNS mergers will not result in an observable GRB. The prompt gamma-ray emission is strongly directional. It has been estimated that only $1-2$\% of KNe detected by gravitational wave observatories might be observable in gamma-rays \citep{metzger2019,colombo2022}.
However, recent work by \citet{dimitrova2023} argues that nearby short GRBs have, on average, broader jets ($\Theta_{\rm jet} \gtrsim \SI{30}{\degree}$) than the more narrowly-beamed cosmological short GRBs due to detection selection effects. 
Determining the fraction of gravitational wave events arising from BNS mergers with associated short GRBs is of key importance. This fraction might be significantly higher than previously expected.

For several long GRBs (e.g., GRB 060605 and GRB 060614), which are typically connected with massive star collapse, deep optical observations excluded an accompanying supernova \citep{Fynbo2006}. Furthermore, GRBs with short peaked gamma-ray emission followed by a spectrally softer extended emission (EE-SGRBs) have been proposed to originate from mergers of compact objects \citep{Norris2002, Norris2006, Gehrels2006}. Follow-up observations of GRB 211211A with a duration of more than $\SI{30}{\second}$ at a distance of $350 \, \text{Mpc}$ revealed a kilonova \citep{Rastinejad2022, Troja2022}, providing a possible solution to the mystery of long GRBs without supernovae. Very recent observations of the exceptionally bright, long-duration GRB 230307A also suggest the presence of a kilonova \citep{Levan_GCN33569,Bulla_GCN33578}. Therefore, follow-up observations of long-duration GRBs with \emph{QUVIK} will be important and some of them may lead to the discovery of new KNe. This highlights the enhanced discovery potential of the mission with an onboard GRB detector.

 \begin{figure}[b!]
    \centering
    \resizebox{0.8\textwidth}{!}{\includegraphics{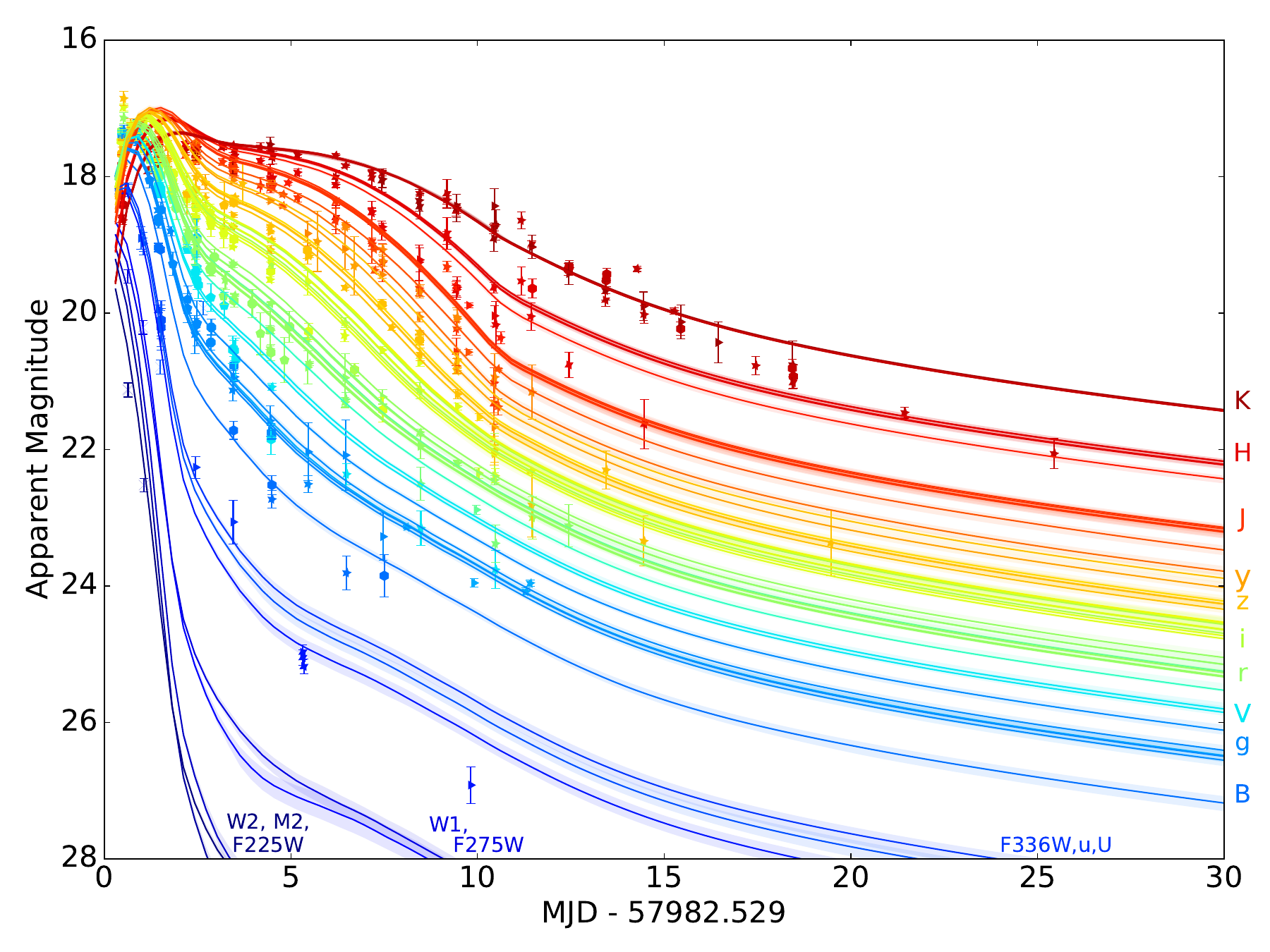}}\vspace*{2mm}
    \caption{The UV–optical–NIR light curves of AT2017gfo and spherically symmetric three-component models from \cite{villar2017}. The data were assembled from 18 different papers and 46 instruments and include 647 individual measurements obtained from $0.45 \, \text{days}$ to $29.4  \, \text{days}$ after the merger. The  three-component model includes a “blue” lanthanide-poor component (opacity component $\kappa = 0.5 \, \text{cm}^2 \, \text{g}^{-1}$) with ejecta mass of $M_{\rm ej} \approx 0.020 \, M_\odot$ and a velocity $v_{\rm ej} \approx 0.27c$; an intermediate opacity component ($\kappa = 3 \, \text{cm}^2 \, \text{g}^{-1}$) with $M_{\rm ej} \approx 0.047 \, M_\odot$ and $v_{\rm ej} \approx 0.15c$; and a “red” lanthanide-rich component ($\kappa = 10 \, \text{cm}^2 \, \text{g}^{-1}$) with $M_{\rm ej} \approx 0.011 \, M_\odot$ and $v_{\rm ej} \approx  0.14c$. The light curves show clearly that while in  NIR and optical the kilonova was visible for weeks its UV emission dropped extremely fast. }%
    \label{fig:KN_lc}
\end{figure}

The optical and UV counterparts of GW170817 were only discovered 11 hours and 15 hours, respectively, after the coalescence of neutron stars \citep{abbott2017a,evans2017}. 
Fig. \ref{fig:KN_lc} from \citet{villar2017} presents UV, optical, and near-infrared (NIR) light curves along with spherically symmetric three-component models calculated for different filters with the highest likelihood scores and their 1$\sigma$ uncertainties. Earlier observations, which are necessary for the understanding of the physics of KNe, remain unavailable. Here, we argue that a UV-photometry mission with a fast-repointing capability, enabling observations of the early emission, would result in real breakthroughs in our understanding of KNe and their nucleosynthesis.

\subsection{Localization of BNS mergers with gravitational-wave detectors}

The estimated rate of mergers involving neutron stars within the distance of 200$\, \text{Mpc}$ ranges from a few to a few tens per year \citep{abbott2020a}. With further upgrades, the sensitivities of the Laser Interferometer Gravitational-Wave Observatory \citep[LIGO;][]{2015CQGra..32g4001L}, Virgo \citep{2015CQGra..32b4001A}, and the Kamioka Gravitational Wave Detector \citep[KAGRA,][]{2021PTEP.2021eA101A} will continue to increase. After 2027 it is expected that a single detector of advanced LIGO will be able to detect BNS mergers up to the distance of 240--325$\, \text{Mpc}$; advanced Virgo alone up to 150--260$ \, \text{Mpc}$; and KAGRA\footnote{\url{https://observing.docs.ligo.org/plan/}} alone up to $128 \, \text{Mpc}$. The full LIGO-Virgo-KAGRA (LVK) GW network is expected to detect BNS mergers up to the redshift $z\approx0.3$ \citep[luminosity distance of $\sim1600 \, \text{Mpc}$;][]{petrov2022}.
Importantly, by mid-2020s, during the observation run O4, the credible region of the localisation area of a significant number of BNS mergers is predicted to reach $\lesssim 100$~square degrees \citep{abbott2020a,petrov2022}. After 2027, during the observation run O5, the number of well-localised BNS mergers (90\% credible area $\le 100$~square degrees) by the full LVK GW network is expected to be 9--90$\, \text{yr}^{-1}$ \citep{petrov2022}.

\subsection{Detectability of KNe with \emph{QUVIK}}

\subsubsection{Luminosity of KNe}
\label{sec:lum-KNe}

 So far, only three kilonovae have been covered well by multi-wavelength observations: GRB 160821B \citep{Lamb2019,Troja2019}, GW170817 / AT2017gfo \citep{abbott2017b}, and GRB 211211A \citep{Rastinejad2022}. Moreover, only a few KN candidates accompanying gamma-ray bursts have been observed: GRB 050709 \citep{Fox2005,Hjorth2005,Jin2016}, GRB 060614 \citep{GalYam2006, Yang2015}, and GRB 130603B \citep{Tanvir2013}. The luminosity distribution of KNe is thus currently not well known and observations provided by \emph{QUVIK} are expected to be of high value.

\begin{figure}[h!]
    \centering
    \resizebox{0.70\textwidth}{!}
    {\includegraphics{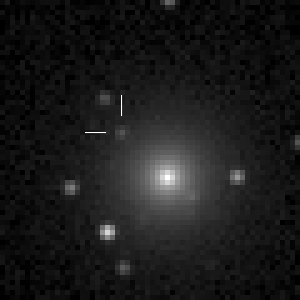}}\vspace*{2mm}
     \caption{A simulated NUV band \emph{QUVIK} image (pixel resolution of 1.75\,arcsec\,px$^{-1}$, FoV cutout of $1.75 \, \text{arcmin} \times 1.75 \, \text{arcmin}$) of a kilonova of $m_{\mathrm{AB}} = 21$ at an angular separation of 1.5 effective radii $(R_{\mathrm{e}})$ from the centre of its host galaxy at a distance of $200 \, \text{Mpc}$ with an exposure time of 1000\,s. The simulation assumed the star field and the host galaxy of AT2017gfo \citep[see][]{2017ApJ...848L..16S}.}
    \label{fig:simulatedKN}
\end{figure}

Figure \ref{fig:simulatedKN} shows a simulated NUV band \emph{QUVIK} image of a $m_{\mathrm{AB}} = 21$ KN at an angular separation from its host galaxy of 1.5 effective radii $(R_{\mathrm{e}})$ at a distance of 200\,Mpc with an exposure time of 1000\,s. We simulated a KN of $m_{\mathrm{AB}} = 21$ for illustrative purpose. Most KNe will be dimmer as seen from Fig.~\ref{fig:KN_sim_lc} and we expect $3.1^{+4.7}_{-2.3}$\,yr$^{-1}$ KNe detectable by \textit{QUVIK} brighter than $m_{\mathrm{AB}} = 21$ to occur at a distance up to 200\,Mpc. \emph{QUVIK}'s angular resolution will enable it to clearly detect a KN on the background of its host, unless it occurs in the galaxy's bright central region.

\begin{figure}[b!]
    \centering
    \resizebox{0.46\textwidth}{!}{\includegraphics{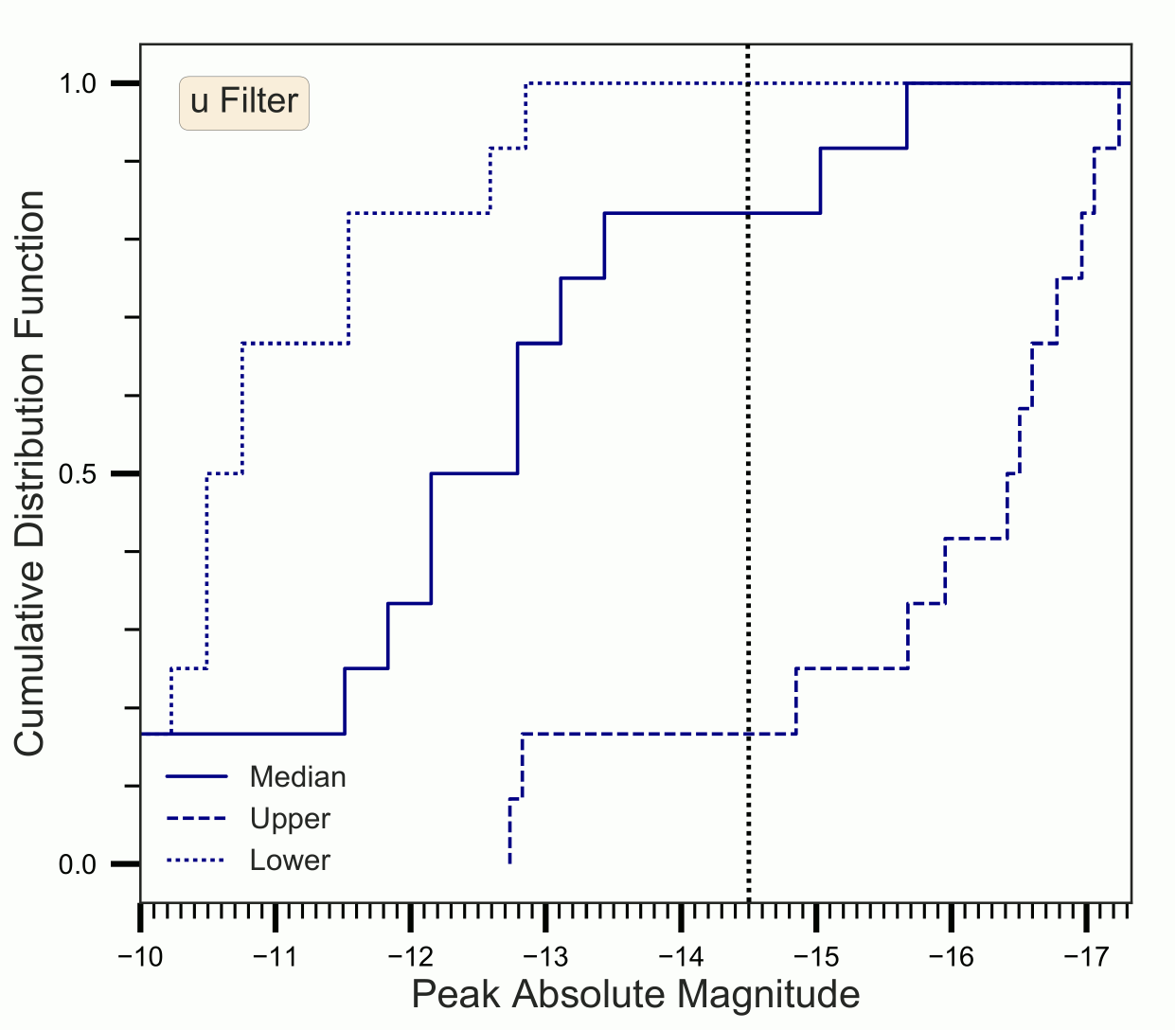}}\vspace*{2mm}
    \resizebox{0.53\textwidth}{!}{\includegraphics{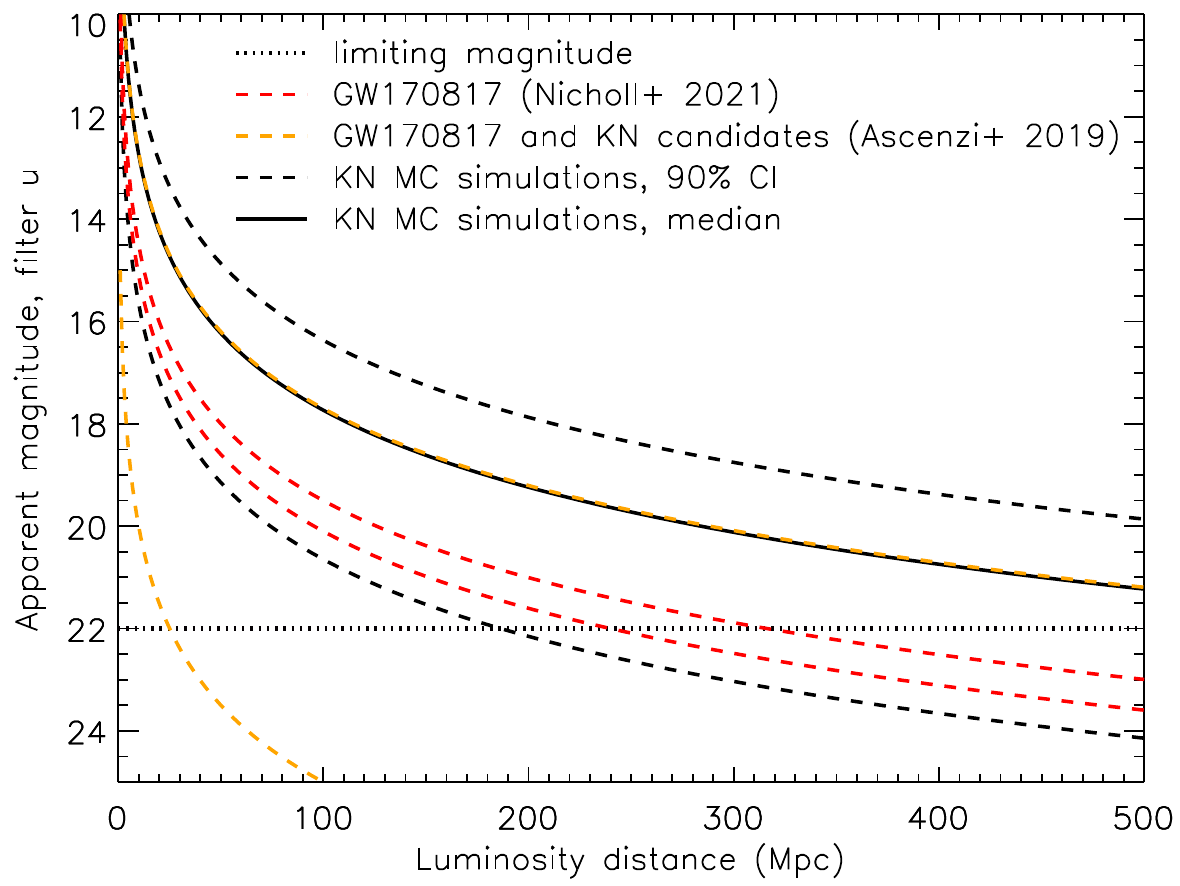}}\vspace*{2mm}
    \caption{Left panel: Cumulative peak luminosity distribution in $u$ filter of kilonova GW170817 / AT2017gfo, KNe candidates GRB 130603B, GRB 050709, GRB 060614 including GRB 150101B, GRB 050724A, GRB 061201, GRB 080905A, GRB 150424A, and GRB 160821B regarded as KNe events, i.e.\ the normalized number of events with peak luminosity lower than the given value. The blue solid, dashed, and dotted lines mark the median, the upper, and the lower limits of the distribution. The vertical black dotted line marks the \emph{QUVIK} limiting AB magnitude of 22 for a KN placed at the distance of 200\,Mpc \cite[adpoted with modification from][]{ascenzi2019}. Right panel: Apparent magnitude of KNe vs distance in $u$ filter. The red curve marks the range of the peak apparent magnitude from kilonova GW170817 / AT2017gfo as modelled by \cite{nicholl2021} in the M. Blanco DECam $u$ filter \citep{flaugher2015} if the KN is placed to a different distance. The yellow curve marks the range of confirmed KNe and KNe candidates using the lower and upper limit of the peak absolute magnitudes shown in the left panel adopted from \cite{ascenzi2019}. The black curve denotes the median and the 90\% CI of the simulated apparent AB magnitudes in the SDSS $u$ filter \citep{york2000} obtained from the nucleosynthesis-powered KNe spectral energy distributions pre-computed by the MC radiative transfer code \texttt{POSSIS 2.0} \citep{anand2023,bulla2023a}.}%
    \label{fig:KN_lum_Ascenzi_mag_dl}
\end{figure}

\begin{figure}[h]
    \centering
    \resizebox{0.49\textwidth}{!}{\includegraphics{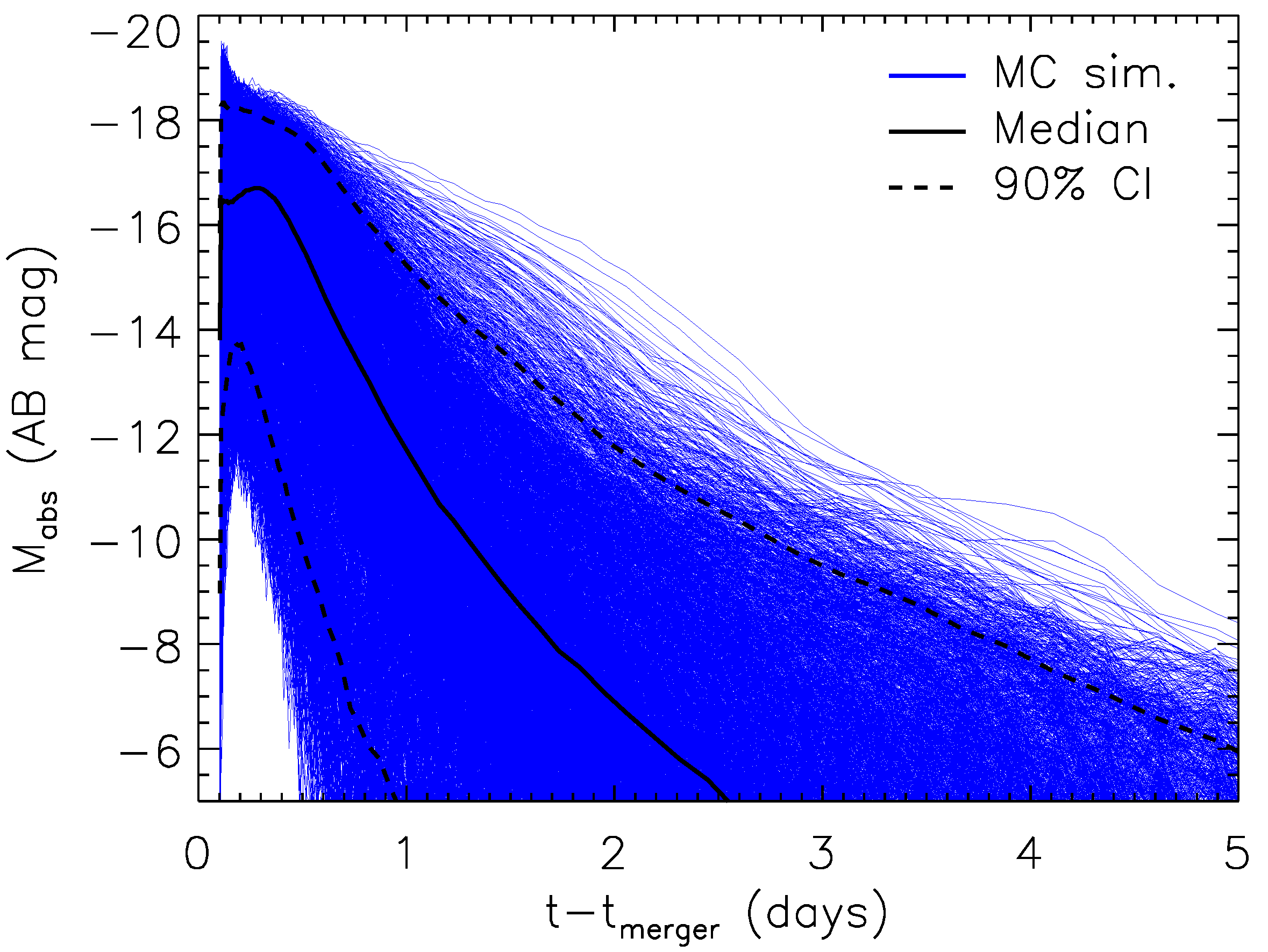}}\vspace*{2mm}
    \resizebox{0.49\textwidth}{!}{\includegraphics{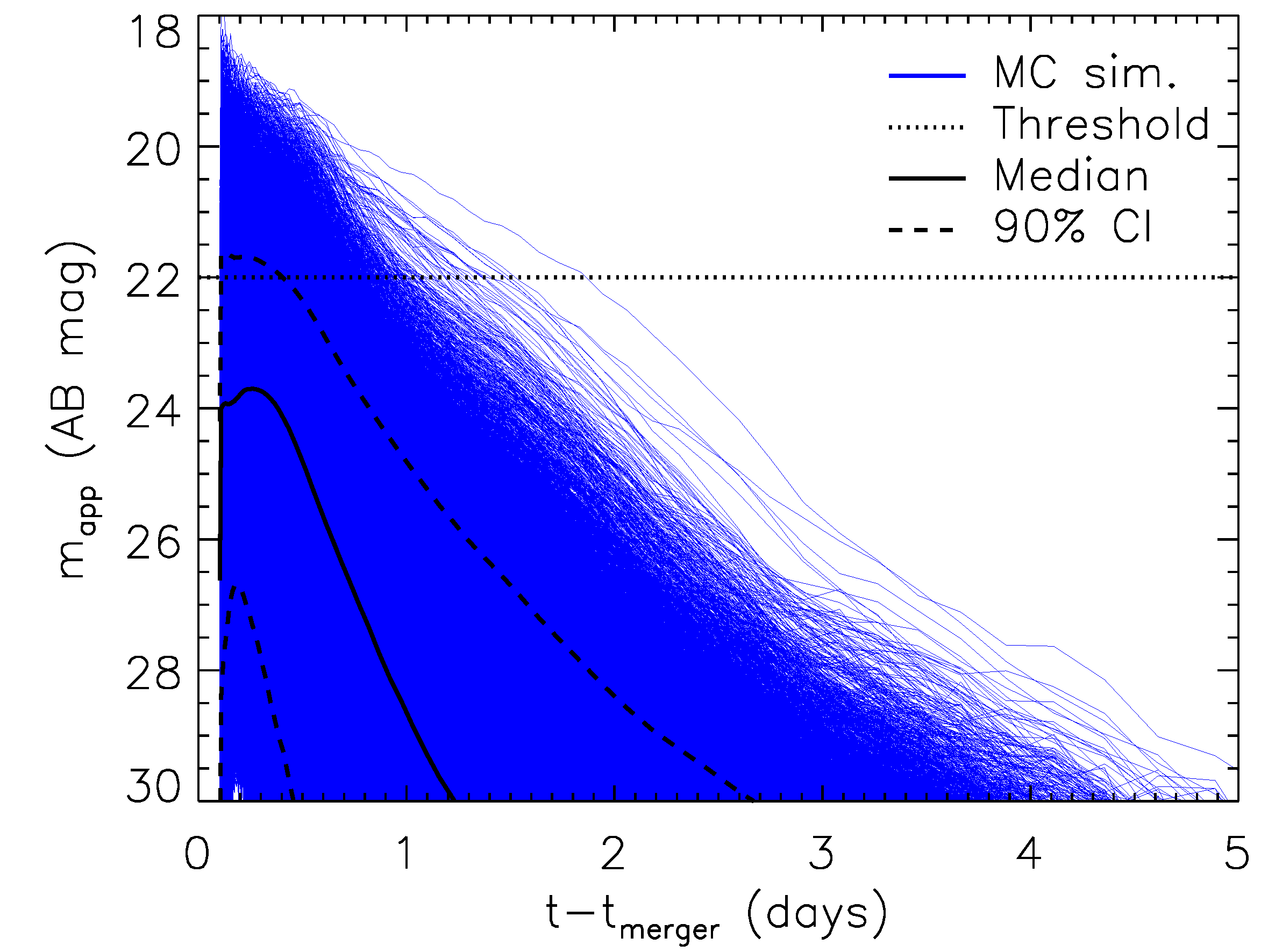}}\vspace*{2mm}
    \caption{Left panel: 
    Around 3\,600 simulated KNe light curves (blue curves) viewed by the \emph{QUVIK} telescope in the 260--360\,nm band in absolute magnitudes in AB magnitude system using the quantum efficiency QE$(\nu)$ of the Gpixel GSENSE4040BSI sensor. Light curves were obtained from the nucleosynthesis-powered KNe spectral energy distributions pre-computed by the \texttt{POSSIS} \citep{bulla2019}.
    3D MC radiative transfer code \texttt{POSSIS 2.0} \citep{anand2023,bulla2023a}.
    No other components, such as heating from the free neutron decay or cocoon emission, were added. Right panel: 
     A sample of about 90\,000 simulated KNe light curves (blue curves) displayed in apparent AB magnitudes randomly drawn from the distribution shown on the left panel and randomly placed (following uniform number density) in the volume up to the distance of 1\,600\,Mpc. A limiting AB magnitude of 22 is marked by the dotted line. The simulations reveal that the fraction of detectable nucleosynthesis-powered KNe for the limiting AB magnitude of 22 (23) is 16\% (44\%).}
    \label{fig:KN_sim_lc}
\end{figure}

The left panel of Fig.~\ref{fig:KN_lum_Ascenzi_mag_dl} from \citet{ascenzi2019} presents the cumulative peak $u$ band luminosity distribution for ten short GRBs modelled with synchrotron afterglow emission plus the contribution of the KN in the UV/optical/NIR wavelengths powered by the radioactive decay of r-process elements. The peak absolute magnitude ranges from $-10.0$ to $-17.3$. Recent work by \cite{sagues2021} presents the distribution of nucleosynthesis-powered KN light curves in the ZTF $gri$ filter system \citep{bellm2019} using the 3D Monte Carlo (MC) radiative transfer code \texttt{POSSIS}\footnote{\url{https://github.com/mbulla/kilonova_models/}} \citep[POlarization Spectral Synthesis In Supernovae;][]{bulla2019}. Time-dependent spectral energy distributions (SEDs) are computed for different viewing angles and used to construct multi-band light curves. 
We employed the new KN model grid \citep{anand2023} of SEDs pre-computed with the improved version of the code, \texttt{POSSIS 2.0} \citep{bulla2023a}. The model describes axially symmetric two-component ejecta (high-velocity lanthanide-rich dynamical ejecta component close to the merger's equatorial plane, lanthanide-poor material closer to the orbital axis, and a disk-wind component at lower velocities) and from these SEDs, we obtained the simulated distribution of the KN light curves in the SDSS $u$ filter \citep{york2000}.
In this improved \texttt{POSSIS 2.0} code, the nuclear heating rates, thermalisation efficiencies, and wavelength-dependent state-of-the-art opacities taken from \citet{tanaka2020} depend on the local properties of the ejecta and time.
The two-component ejecta model is simulated for 11 different viewing angles $\Theta_\mathrm{obs}$ (equally spaced in cos($\Theta_\mathrm{obs}$) from a face-on/jet axis to the edge-on/merger plane) and is characterised by five parameters: the mass of the dynamical ejecta $M_\mathrm{ej,dyn}=[0.001,0.005,0.010]M_{\odot}$, the averaged velocity of the dynamical ejecta $\bar{\nu}_\mathrm{ej,dyn}=[0.15,0.20,0.25]c$, the averaged electron fraction of the dynamical ejecta $\bar{Y}_\mathrm{e,dyn}=[0.15,0.20,0.25]$, the mass of the disk-wind ejecta $M_\mathrm{ej,wind}=[0.01,0.05,0.09,0.13]M_{\odot}$, and the averaged velocity of the disk-wind ejecta $\bar{\nu}_\mathrm{ej,wind}=[0.05,0.10,0.15]c$ \citep[for details see][]{anand2023}.
The median absolute peak AB magnitude is $-17.4$ (90\% confidence interval is between $-14.9$ and $-18.6$~mag).

Note that the opacities from \citet{tanaka2020} are computed up to the ionisation stage of IV, which is valid for ejecta temperatures roughly below 20\,000\,K. At early times (earlier than about 0.5--1\,days after the merger), the temperatures in the ejecta are higher and the material can be ionised to higher stages. The lack of opacity contribution from elements in higher ionisation stages in the modelling might result in an overestimated brightness in the early UV light curves \citep{bulla2023a} by $\sim 1 \, \text{mag}$ when compared to models including contributions up to ionisation stage XI \citep{Banerjee2023}.
Note, however, that this model does not include possible additional brightening from the free neutron beta decay and the shock-cooling of the material surrounding the merger remnant, which, as discussed in Sect. \ref{constraints}, might increase the early UV luminosity of KNe by as much as 2 magnitudes. 

Having estimates for the peak absolute magnitude distribution of the KN emission, one can plot the apparent magnitudes for different distances and compare them with a given detection threshold. This is presented in the right panel of Fig.~\ref{fig:KN_lum_Ascenzi_mag_dl}.

The left panel of Fig.~\ref{fig:KN_sim_lc} shows a similar simulation of nucleosynthesis-powered KN light curves with the \texttt{POSSIS 2.0} code for the 260--360~nm band assuming the preliminary mean optical throughput of the
telescope in the NUV band (the exact coatings and filters and hence the throughput of the telescope as a function of wavelength will be determined in the future development phases) and the quantum efficiency QE$(\nu)$ function of the Gpixel GSENSE4040BSI sensor, which is a candidate for \emph{QUVIK}'s NUV detector (see Sect. \ref{sect.payload}). For this band and sensor, the median absolute peak AB magnitude is 
$-17.3$, and the 90\% CI is between $-14.4$ and $-18.6$.

\subsubsection{Detectable kilonovae rate}

There is considerable uncertainty in the estimated BNS merger rate in the local Universe. Recent estimate of BNS coalescence rate density is $320^{+490}_{-240}$~Gpc$^{-3}$~yr$^{-1}$ \citep{abbott2021}, which corresponds to a coalescence rate of $11^{+16}_{-8}$~yr$^{-1}$ up to a distance of 200\,Mpc.
The BNS merger detection rate expected during the fifth observing run (O5) of the full LVK GW network is $190^{+410}_{-130}$~yr$^{-1}$ \citep[90\% credible interval,][]{petrov2022}.
To provide 
an estimate for the number of KNe detectable by \emph{QUVIK}, we assume purely nucleosynthesis-powered KNe and the predictions for BNS mergers observed by the full LVK GW network in the O5 observing run. We simulated $\sim100$ million KNe uniformly distributed in a sphere with a radius of 1\,600\,Mpc (viewing angles were equally spaced in cosine and the ejecta model parameters were sampled as described in Sec.~\ref{sec:lum-KNe}). Then, we used the distribution of $\sim3\,600$ absolute peak AB magnitudes of simulated KNe with the \texttt{POSSIS 2.0} code as shown in Fig.~\ref{fig:KN_sim_lc} to randomly assign to each of the $\sim100$ million KNe an absolute peak AB magnitude. Next, for a given limiting magnitude threshold, we calculated the fraction of simulated KNe $f_\mathrm{det.EM}$ for which the EM counterpart would be detectable by \emph{QUVIK}. The simulations were performed in the 260--360~nm band employing the quantum efficiency QE$(\nu)$ of the Gpixel GSENSE4040BSI sensor. Next, we assumed the BNS merger detection rate foreseen for the
O5 run by the full LVK GW network 
$190^{+410}_{-130}$~yr$^{-1}$ (90\% credible interval) as mentioned above by \cite{petrov2022}.
One has to consider that only a fraction $f_\mathrm{BNS,KN}$ of BNS mergers will produce a KN; for the remaining fraction, the merger will result in a prompt black hole collapse without a disk or ejecta. We adopt the value of $f_\mathrm{BNS/KN}=0.78$ \citep{colombo2022}. The last two factors are the duty cycle $f_\mathrm{d}$ of the telescope and the fraction of the sky visible by \emph{QUVIK}. We assume a 50\% duty cycle. As for the sky visibility fraction, we assume $f_\mathrm{sky}=0.8$. Thus, the predicted number of detectable KNe is calculated as $N_\mathrm{KN}(\mathrm{yr}^{-1})=N_\mathrm{BNS}(\mathrm{yr}^{-1})~f_\mathrm{det.EM}~f_\mathrm{BNS/KN}~f_\mathrm{d}~f_\mathrm{sky}$. 

Our simulations reveal that the fraction of detectable nucleosynthesis-powered KNe for the limiting AB magnitude of 22 is 16\% in the 260--360\,nm band.
This fraction increases to 44\% if we target the limiting AB magnitude of 23, which will require longer exposures and at least two orbits with \emph{QUVIK} (see Sect. \ref{sect.payload}).
The right panel of Fig.~\ref{fig:KN_sim_lc}, displays 
$\sim3600$ of simulated nucleosynthesis-powered KNe light curves in apparent AB magnitudes randomly distributed in a volume up to the distance of 
1600\,Mpc. The plot indicates that more efficient probing of nucleosynthesis-powered kilonova light curves will require deeper multi-orbit exposures, targeting the AB magnitude of 23.
We estimate that for the limiting AB magnitude of 22, the expected rate of detectable KNe is $9.3^{+20}_{-6.4}$\,yr$^{-1}$. Similarly, by using the BNS coalescence rate of $11^{+16}_{-8}$~yr$^{-1}$, as mentioned at the beginning of this subsection, we expect the rate of detectable KNe up to the distance of 200\,Mpc to be 
$3.2^{+4.9}_{-2.4}$\,yr$^{-1}$ ($3.1^{+4.7}_{-2.3}$\,yr$^{-1}$) for the limiting AB magnitude of 22 (21).
The estimates are summarised in Table~\ref{tab:detectable_rate}, which provides the expected detectable KNe rates for a number of limiting magnitudes. 

Note that the simulated UV light curves may be overestimated in brightness by $\sim1$~mag in the early time as mentioned above; however, also note that the light curves did not account for the additional emission from the free neutron decay, which may increase the NUV luminosity during the first few hours by a magnitude or more \citep{metzger2019}. The simulations did also not include possible additional brightening from the shock-cooling of the material surrounding the merger remnant, which can be an important source of early emission in NUV and FUV. The expected constraints on these additional emission components are discussed in the next section. Given that no early emission of a KN has been observed since GW170817, every observation by \emph{QUVIK} will be highly valuable.

\begin{table}[h]
\centering
\begin{tabular}{cc}
  \hline
  \hline
  Limiting     &        Rate   \\
  AB magnitude & (yr$^{-1}$) \\
  \hline
  \hline\\[-2ex]
  21.0 & $2.4^{+5.2}_{-1.6}$     \\[1ex]
  21.5 & $4.8^{+10}_{-3.3}$    \\[1ex]
  22.0 & $9.3^{+20}_{-6.4}$   \\[1ex]
  22.5 & $16^{+35}_{-11}$   \\[1ex]
  23.0 & $26^{+57}_{-18}$  \\[1ex] 
  \hline\\
\end{tabular}
\caption{Predicted detectable KNe rate by \emph{QUVIK} in the 260--360\,nm band following GW events from BNS mergers during the LVK O5 observing run \citep{petrov2022} for different limiting magnitudes. }
\label{tab:detectable_rate}
\end{table}

\subsection{Constraints from UV light curves of kilonovae
\label{constraints}}

While the red and infrared emission components of the kilonova AT2017gfo were most likely produced by the radioactive decay of newly produced r-process nuclei, there is a debate about the observed UV and blue emission, which was dominant in the first $\sim$1.5\,days. The observed early blue emission is well-explained by radioactive material with a relatively low opacity \citep["blue" KN;][]{metzger2010}. However, the implied large quantity of low-opacity fast material is not predicted by the simulations of dynamical or disk ejecta. A possible explanation is that the merger did not immediately result in a black hole but produced a short-lived magnetar, which is responsible for the large amounts of fast ejecta with a high electron fraction \citep{metzger2018}. This has recently also been supported by general relativistic magnetohydrodynamic numerical simulations \citep[e.g.][]{curtis2023,combi2023}. Very early UV emission could also arise from shock interaction, so-called cocoon emission \citep{Nakar2017}, or the beta decay of free neutrons \citep{kulkarni2005}. UV observations obtained early after the neutron star coalescence will be able to distinguish between the various models.

\citet{Piro2018} suggested that the power-law evolution of the luminosity during the early time of AT2017gfo could be explained by the cooling of shock-heated material around the neutron star merger. This heating could be the result of the interaction of the gamma-ray burst jet with the merger debris, the so-called cocoon emission theoretically predicted by \citet{Nakar2017} and \citet{Gottlieb2018}. Possible additional brightening of a KN from the shock-cooling of the material surrounding the merger remnant can be an important source of emission in NUV and FUV. The peak luminosity can be higher by 2\,mag \citep{Kulkarni2021} compared to the purely nucleosynthesis-powered KN and would be important in the first several hours after the merger. It was suggested that the cocoon emission is likely the first UV/optical signal that can be seen when the jet responsible for the short GRB, which follows the BNS merger, is viewed off-axis \citep{Kulkarni2021}.

\begin{figure}[h]
    \centering
    \resizebox{0.49\textwidth}{!}{\includegraphics{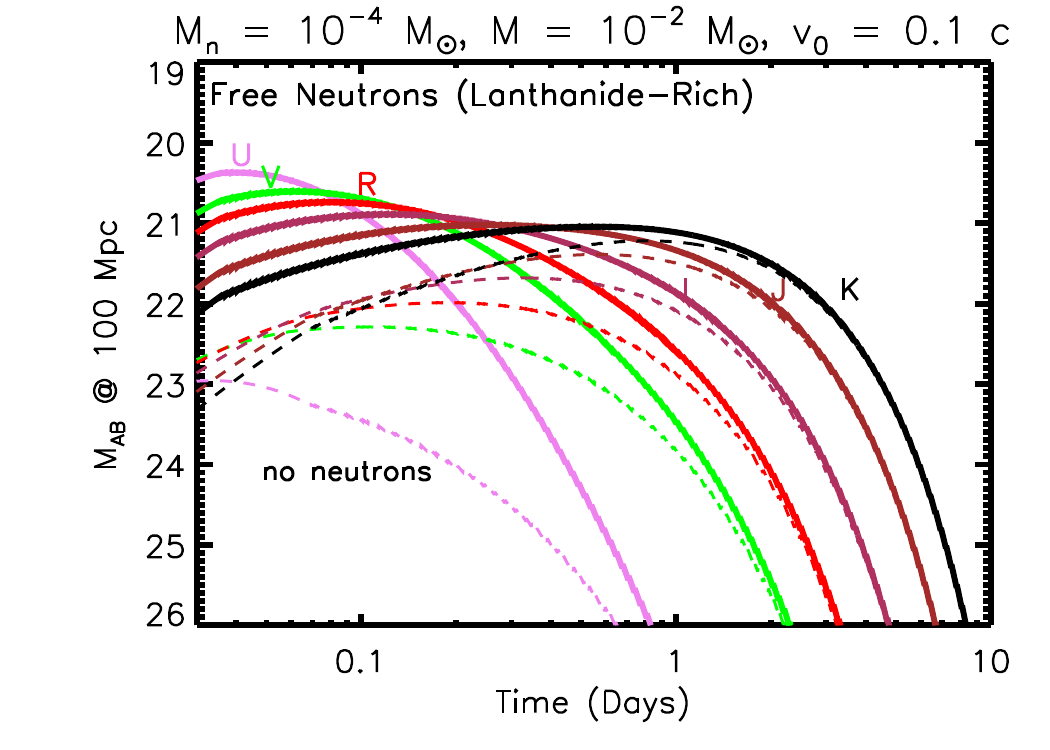}}\vspace*{2mm}
    \resizebox{0.49\textwidth}{!}{\includegraphics{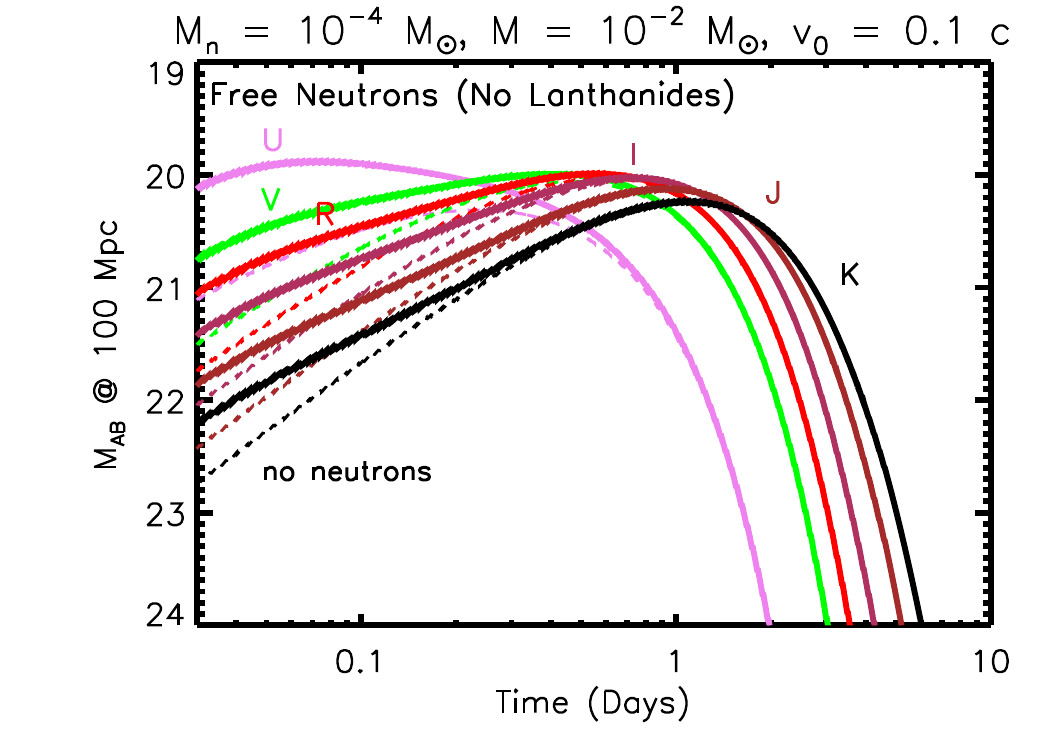}}\vspace*{2mm}
    \caption{Left panel: Model of a red KN (lanthanide-rich) including the emission from free neutron decay (``neutron precursor'' emission, solid curve) from the outer layers of the ejecta containing the mass of neutrons $M_{\rm n} = 10^{-4}\,M_{\odot}$.  Right panel: Similar to the left panel, but for a blue KN (lanthanide-poor).  Both models were calculated for the same total ejecta mass $M = 10^{-2}\,M_{\odot}$ and velocity $v_0 = 0.1 c$.  For comparison, the dashed curves are the models without free neutron decay emission. From \citet{metzger2019}.}%
    \label{fig:kn_free_neutrons}
\end{figure}

Free neutrons, if present in the outer ejecta layers, would decay as $n^0 \rightarrow p^+ + W^- \rightarrow p^+ + e^- + \bar{\nu}_e$ with a half-life of $\sim$15\,minutes, providing extra heating and enhancing the early KN emission \citep{kulkarni2005,metzger2015} which, as indicated in Fig.~\ref{fig:kn_free_neutrons}, would increase the NUV luminosity during the first few hours by a magnitude or more \citep{metzger2019}. In the FUV band, the free neutron decay would increase the luminosity of KNe by as much as two magnitudes \citep{Kulkarni2021}. The predicted free neutron decay in a KN explosion has not yet been observed and early photometry by \emph{QUVIK}, performed less than about 6\,hours after the merger, would thus provide critically important data. 

Early observations in NUV and FUV with \emph{QUVIK}, in the first few hours after the BNS merger, will be crucial for distinguishing between models. \citet{dorsman2023} performed a Bayesian analysis to determine whether a UV satellite could distinguish between the physical processes driving the early blue emission component. They show that if the satellite starts collecting data early, within a couple of hours, it will be able to distinguish between early radiation models. In particular, probing the beta decay of free neutrons requires light curves taken less than 6\,hours after the BNS merger \citep[see also][]{Kulkarni2021}. \citet{dorsman2023} also show that having simultaneously taken UV and optical data improves the constraints on models significantly. Given the planned extensive ground-based follow-up efforts, the availability of complementary optical and NIR data is likely.

\begin{figure}[h]
    \centering
    \resizebox{0.8\textwidth}{!}{\includegraphics{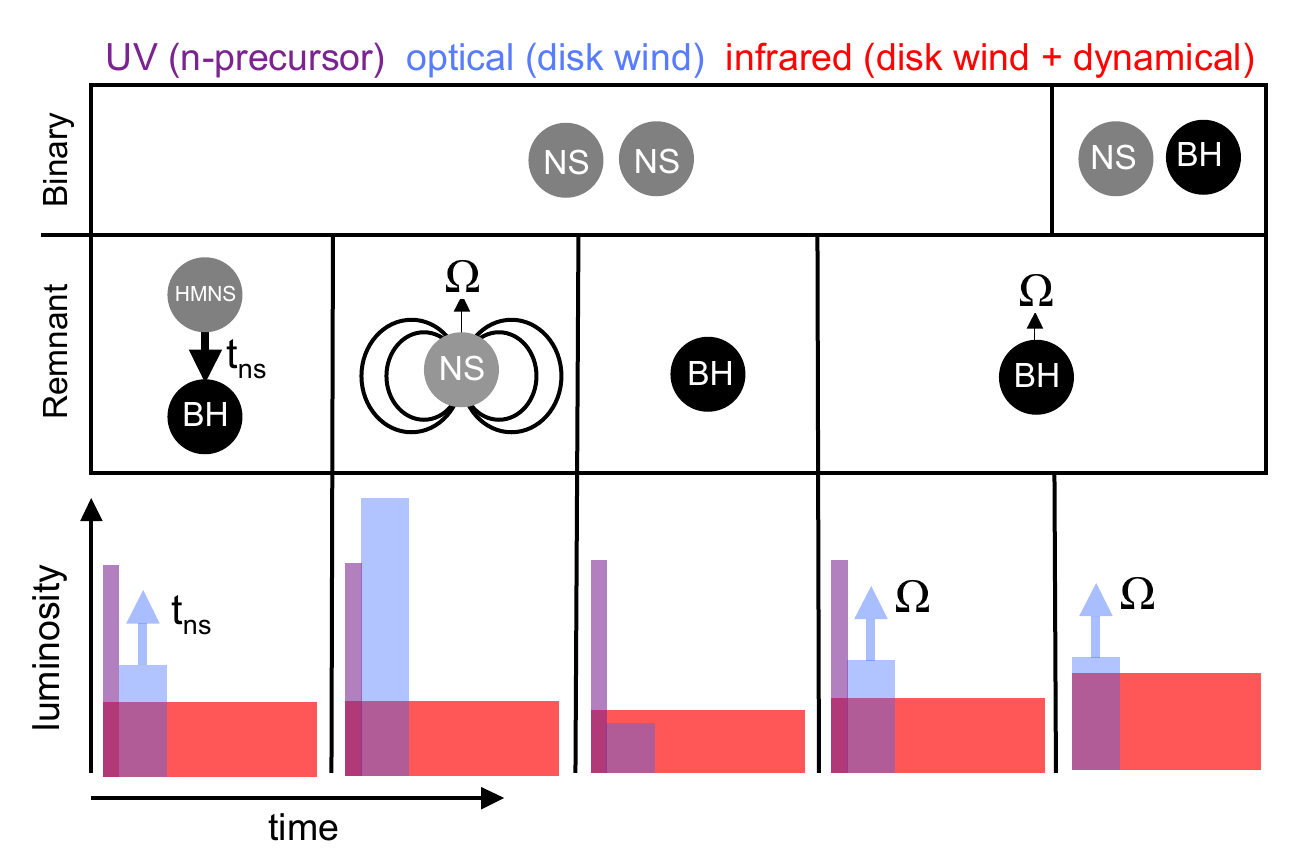}}\vspace*{2mm}
    \caption{Illustration of various scenarios of the NS-NS or NS-BH merger (indicated in the top panel), the resultant remnant (middle panel) and the produced relative amount of UV/blue emission from a neutron precursor (purple), optical emission from the lanthanide-free material (blue) and IR emission from the lanthanide-rich ejecta (red). $t_{\rm NS}$ denotes the time after which the hypermassive neutron star (HMNS) merger remnant collapses to a black hole. The Greek letter $\Omega$ denotes a spinning magnetized neutron star and a rapidly spinning black hole remnant \citep[from][]{kasen2015}.}
    \label{fig:outcomes}
\end{figure}

The ratios between the KN fluxes obtained in the UV, optical, and NIR bands will allow us to identify and constrain the properties of the different ejecta \citep{metzger2019}. In particular, the UV observations performed in the first few hours of the KN will probe the fastest ejecta's mass, composition, and thermal content and allow us to constrain its geometry, quantity, and kinematics. Fig. \ref{fig:outcomes} shows that early multi-wavelength observations will also allow us to determine the nature of the merger product, which can be a hyper-massive neutron star that quickly collapses into a black hole or a stable, rapidly spinning, highly magnetised neutron star. Alternatively, a BNS merger can result in a direct collapse into a black hole. Each of these outcomes produces different ratios of observed UV, optical, and NIR fluxes after the merger \citep{kasen2015,fernandez2016} and if a long-lived stable magnetar is formed, the KN signal might be drastically different \citep[e.g.][]{bucciantini2012,yu2013,metzger2014}. Importantly, since the gravitational wave observations allow us to obtain accurate measurements of the mass of the binary, by ascertaining the outcome of the merger from the early electromagnetic signal, we can, in principle, constrain the equation of state of neutron stars \citep[e.g.][]{bauswein2013,margalit2019}.

\section{Gamma-ray bursts}
\label{sec:grb}

\subsection{GRB physics}

Gamma-ray bursts \citep{Piran2004,Zhang2004,Meszaros2006,Kumar2015} are some of the most luminous explosions in the Universe. They are traditionally divided into two classes, defined by their observed duration in soft $\gamma$-rays: \emph{long GRBs} with T$_{90}>$2\,s and \emph{short GRBs} with T$_{90}<$2\,s, where T$_{90}$ is the time in which 90\% of the soft $\gamma$-rays are emitted. Although this division is somewhat arbitrary, the two classes are indeed connected to two different progenitor systems: Most long GRBs are due to the gravitational collapse of stars with masses reaching tens of $M_\odot$ and are accompanied by broad-line Ic SNe \citep[e.g.][]{Hjorth03, Cano17}. Short GRBs, as has been discussed in the previous section (Sect. \ref{sec:kilonovae}), originate from the merger of neutron stars and are followed by the KN emission. However, there are also cases of merger progenitors for seemingly long GRBs \citep{Rastinejad2022} and collapsar progenitors for short GRBs \citep{Ahumada2021,Rossi22}.

\begin{figure}[h]
    \renewcommand{\familydefault}{\sfdefault}\normalfont
    \centering
 \includegraphics[width=0.75\textwidth]{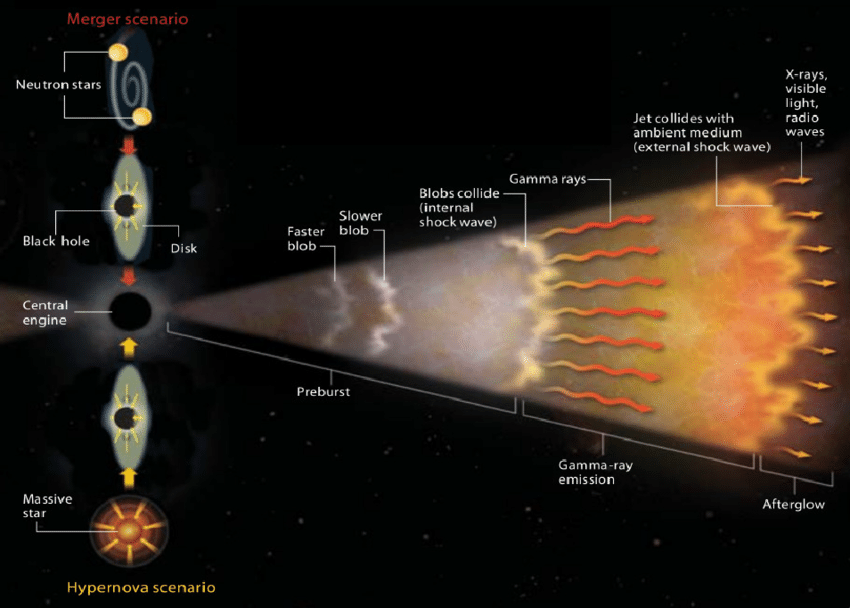}
        \caption{Sketch of the processes leading to a GRB, its prompt emission and afterglow \citep[from][]{Gehrels2002}.}
    \label{fig:sketch}
\end{figure}

\begin{figure}[h]
    \renewcommand{\familydefault}{\sfdefault}\normalfont
    \centering    \includegraphics[width=0.75\textwidth]{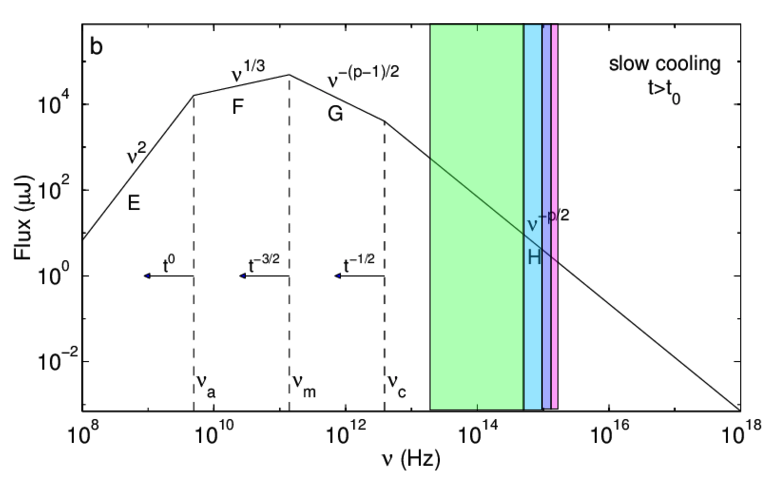}
    \caption{Model of a GRB afterglow in the slow cooling regime and four observing spectral regions: purple denoting the \emph{QUVIK} FUV band; blue and light blue indicating the \emph{QUVIK} and \emph{ULTRASAT} NUV bands, and green for the range 400--1000\,nm observable by ground-based telescopes. Based on \citet{Sari98}.}%
    \label{fig:afterglow}
\end{figure}

The progenitors of both short and long GRBs produce well-collimated ultra-relativistic jets, where collisions of shells with different Lorentz factors are responsible for the so-called ``prompt'' $\gamma$-ray emission (see Fig. \ref{fig:sketch}). When the jet interacts with the interstellar medium, it slows down and produces the \emph{GRB afterglow} emission detectable in a wide range of frequencies from radio to high-energy $\gamma$-rays. The afterglow has a simple synchrotron spectrum characterised by a three-fold broken power law that breaks at characteristic frequencies: at the self-absorption frequency $\nu_a$, the typical frequency $\nu_i$ and the cooling frequency $\nu_c$ \citep{Sari98}.  The later afterglow is located in the ``slow cooling regime'', where $\nu_i < \nu_c$ and the NIR-optical-UV bands are usually located on the same slope above or below the cooling frequency. Fig. \ref{fig:afterglow} shows a schematic spectrum of a GRB afterglow and the location of different wavebands, including the bands envisioned for \emph{QUVIK} and \emph{ULTRASAT}. 

The study of slopes, break frequencies, and their evolution depends on the micro- and macro-physics of the explosion and its environment. These simple power-law spectra are altered by the material within the host galaxy and the material between the host and observer. In particular, dust extinction can introduce a curvature in the spectrum and significantly attenuate the UV emission. 

Long GRB afterglows are on average more luminous than those of short GRBs \citep{Kann11}.  While short GRBs have only been detected up to a redshift of 2.2 \citep{Selsing18}, long GRBs have been found out to $z\sim9.4$ \citep{Cucchiara11}. However, above $z\sim2$ they are no longer observable in UV due to the Ly$\alpha$ dropout, hence GRB science in the UV band focuses predominantly on GRBs below redshift $\sim 2$. 

\subsection{The science potential of UV observations}

\subsubsection{UV data of GRB afterglows}

So far, essentially all UV data of GRB afterglows have come from the \emph{Swift} satellite \citep{gehrels2004}. \emph{Swift} is equipped with the Burst Alert Telescope \citep[BAT;][]{Barthelmy2005}, X-Ray Telescope \citep[XRT;][]{Burrows2005}, and the UV and Optical Telescope \citep[UVOT;][]{roming2005}.  UVOT is a 30\,cm telescope with six different filters ranging from 170 to 600\,nm (uvw2, uvm2, uvw1, u, b, v). When a GRB is detected by BAT, the satellite slews so that the burst gets in the FoV of XRT and UVOT\@. The nominal slewing time is $\sim 100$\,s and most GRBs are observed within the first 150\,s, but the reaction time can be as fast as 40\,s \citep{Roming09}. UVOT is also prepared to receive GRB triggers from other missions and initiate follow-up. Other UV observatories in orbit are UVIT on \textit{AstroSat} and WFC3/STIS on \textit{HST} \citep{Woodgate1998,Kimble2008}. However, their fast response capability is limited.

GRB observations by UVOT have been published in two catalogues: \cite{Roming09} and \cite{Roming2017} which include data between 2004 and 2010.  About half of the observations show a simple power-law behaviour for the temporal slope, but $\sim$ 20\% show clear flares or unusual deviations from a (broken) power-law. \cite{Oates12} found a correlation between the peak magnitude in UV and the temporal decay slope where GRBs with higher peak luminosities decay faster than those with lower luminosities.  The most likely interpretation is that this is not an intrinsic property, but a viewing angle effect, since a similar correlation is also observed in X-rays \citep{Oates15}.  UVOT observations have been an integral part of the GRB afterglow data for getting very early observations and contributing to the broad-band spectral energy distributions across the electromagnetic spectrum.

\subsubsection{Early GRB data: flares and the onset of the afterglow probing jet physics}

UVOT has been vital for many years to observe GRBs at very early times, sometimes even during the prompt emission, observing the rise to maximum light and subsequent brightness decay. In recent years, rapidly slewing robotic optical telescopes have also been able to observe the transition between the prompt and the afterglow emission \citep{Ror2023,Greiner2009,Rykoff2009,Racusin2008,Vestrand2005}.

\begin{figure}[h!]
    \renewcommand{\familydefault}{\sfdefault}\normalfont
    \centering
    \includegraphics[width=0.49\textwidth]{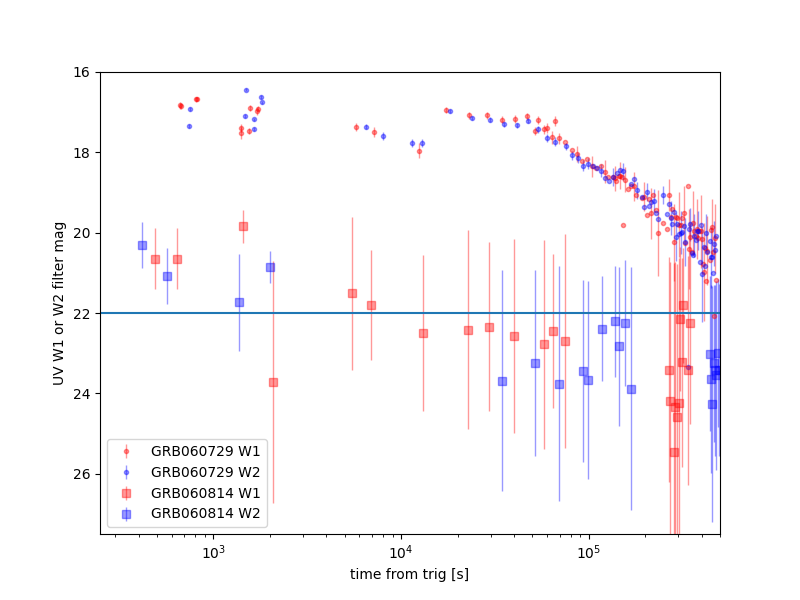}
    \includegraphics[width=0.49\textwidth]{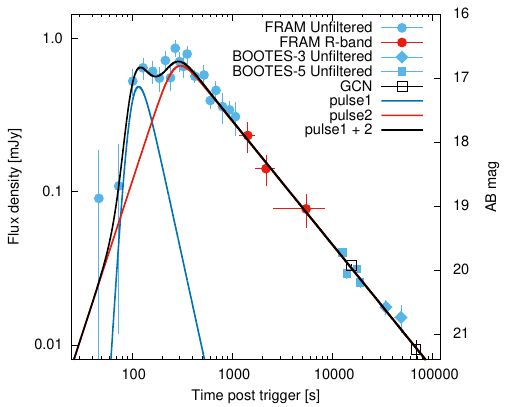}
    \caption{Left: UVOT light curves in two filters for two representative GRBs on the bright and faint ends of the distribution \citep[][data taken from the second UVOT catalogue]{Roming2017}.  Right: Light curve of GRB 190819 showing two prominent flares at early times  \citep[from][]{Jelinek22}.}
    \label{fig:LCs}
\end{figure}

At early times, the afterglow evolution often differs from a simple power-law decay showing features such as bright flares, which can be brighter than the prompt emission, breaks in the decay, or rebrightenings at later times (see also Fig. \ref{fig:LCs}).  Flares have been observed in X-rays \citep{Margutti11, Yi2016} and UV by UVOT \citep{Zaninoni2013, Swenson2013, Yi2017}.  Their origin is still not fully understood; models include a revived central engine activity \citep{Burrows05}, interactions with the interstellar medium (ISM) which might be stratified from previous mass ejections of the progenitor star \citep{Ayache2020}, or reverse shocks that are created when the shock front hits the ISM \citep{Lamberts17}\@. These processes might all be acting at different times, e.g.\ ISM interaction might explain later flaring, while the central engine activity might cause the earlier flares.

Thanks to modern CMOS BSI sensors, \emph{QUVIK} will have better sensitivity than UVOT and it will also observe in the FUV (see Sect. \ref{sect:FUV}). 
This will allow us to sample the temporal evolution of the colour index of flares, brakes, and rebrightenings. The colour index allows to discriminate between different models for the origin of flares from either central engine activity or the ambient medium \citep{Gao2013}. We will also be able to build up a larger sample of features at different times in simultaneous bands to probe the different scenarios. The understanding of the physics of this emission requires close coordination with longer wavelength observations in the optical and NIR. Ground-based facilities such as GROND at the 2.2\,m telescope in La Silla, a 7-channel simultaneous camera \citep{GreinerGROND}, have proven extremely important to obtain dense, simultaneous light curves and thus probe the evolution across different bands \citep{Greiner19}.  

Only a subset of models is able to explain colour changes (variations of colour index): mostly those employing central engine variability. A case of XRF 071031 is discussed in detail in \citet{Kruhler09}.

Observing the very early phases of GRBs, possibly including flares related to the prompt emission, would be extremely beneficial. \emph{ULTRASAT}, covering $\sim0.5$\% of the entire sky in a single pointing, may detect one GRB per year directly within its FoV just by coincidence. Due to its smaller FoV, accidental detection by \emph{QUVIK} is unlikely. Still, an early simultaneous two-band observation of a GRB afterglow will benefit greatly from the presence of the onboard GRB detector with localisation capabilities (see Sect. \ref{sect:GALI}), which will reduce the reaction time to a few minutes. Furthermore, \emph{QUVIK} will also have the ability to perform rapid follow-up observations of GRBs detected and localised by other observatories. 

\subsubsection{Redshift estimates}

Astrophysical sources get absorbed below the Lyman limit at 912\AA{}, the shortest wavelength of the Lyman forest at which the electron of a hydrogen atom in the ground state gets ionised.  This can be used to determine photometric redshifts, as with increasing redshift blue bands subsequently do not detect the object, a technique also called ``Lyman dropout'' used in a search for high redshift galaxies. This has also frequently been used for GRBs, in particular for fast UVOT detections, giving a rough estimate of the redshift very early on.

Considering the current baseline design with the NUV band starting at 260 nm, galaxies at z~$>$~1.8 would show a loss in flux. A non-detection with \emph{QUVIK} could imply z~$>$~3.3. To determine photometric redshifts, usually more bands are required \citep[see e.g.][]{Kruehler11}; ideally, one needs a further band in the optical/NIR range to derive a secure photometric redshift, because a decrease in flux might also be due to extinction from the host galaxy. This is observed to occur for faint bursts at low redshifts. For bright bursts, with a high S/N, a dropout in one band results in a sharp feature, which cannot be mimicked by dust. Redshift
estimates will clearly benefit from 2-band observations. 

\subsubsection{Host galaxy observations}
UV observations of GRB host galaxies are still scarce. To date, only a few GRB hosts have been observed in UV bands using either \textit{Swift}/UVOT, \textit{AstroSat}/UVIT, or \textit{HST}. Massive stars emit a considerable part of their light in the UV, in contrast to older and redder stars and hence directly probe the star-formation rate (SFR) in galaxies.  Particularly for long GRB hosts as star-forming galaxies, UV data are one of the methods to determine the unobscured ongoing SFR. The corresponding relation between UV flux in the 150--280 nm rest frame and the unobscured SFR is described in \cite{Kennicutt98}. 

\begin{figure}[h!]
    \renewcommand{\familydefault}{\sfdefault}\normalfont
    \centering
    \includegraphics[width=1.0\textwidth]{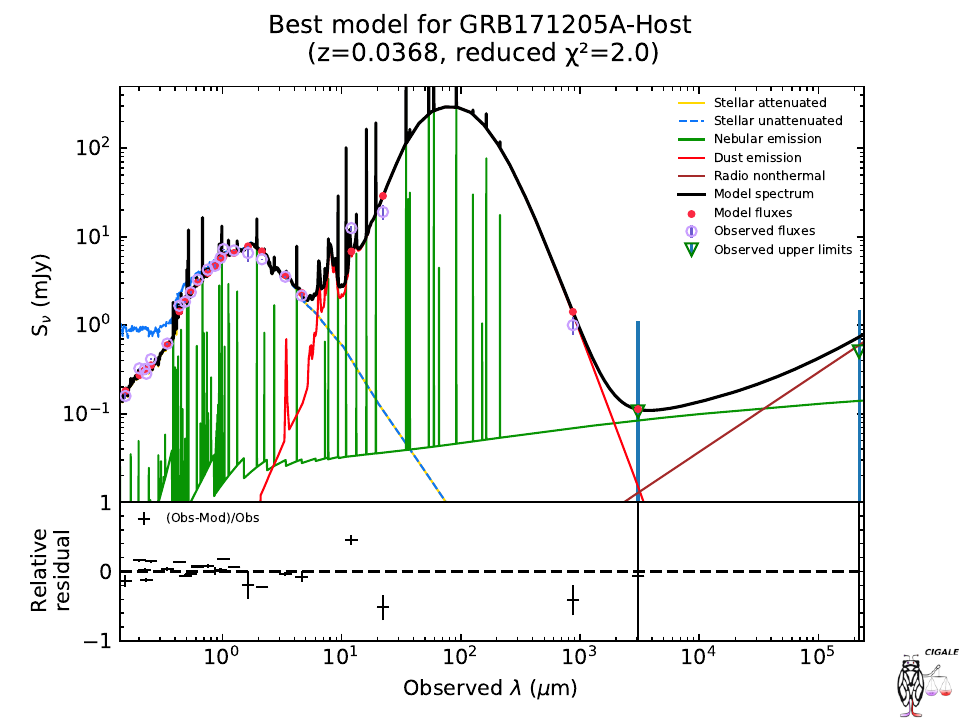}\\
    \includegraphics[width=1.0\textwidth]{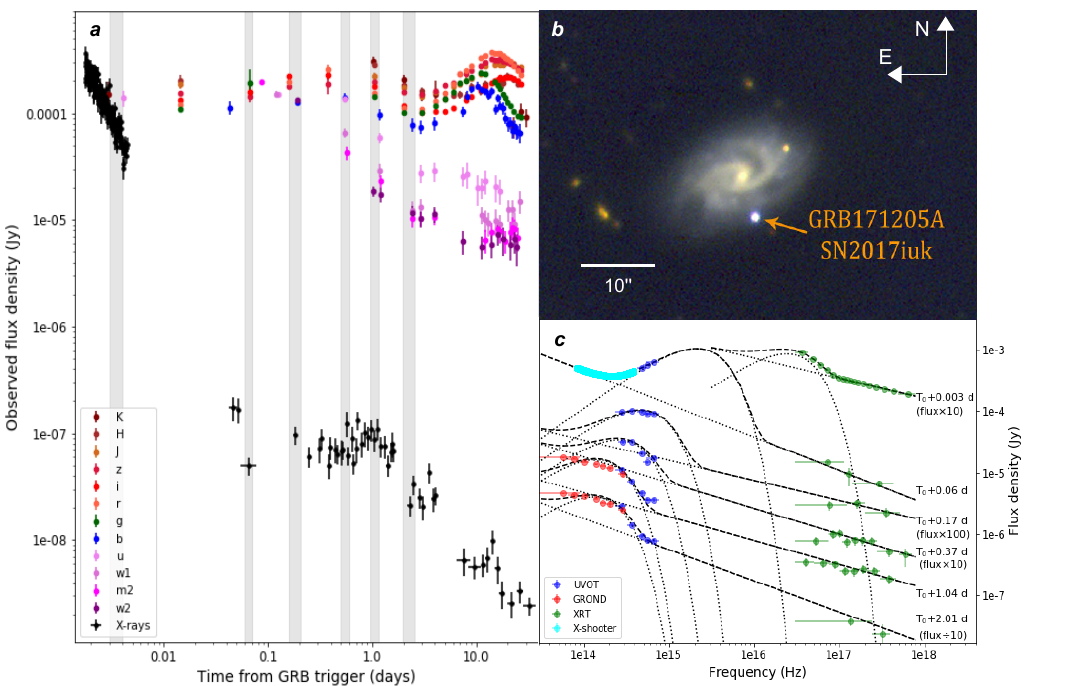}
\caption{Top: SED fit of the host of GRB 171205A (distance 163\,Mpc) including observed UV data from HST at the blue end (from de Ugarte Postigo et al. subm.). Bottom: (a) Evolution of the multi-wavelength light curve of the transient following GRB 171205A. (b) Color image of the host galaxy of GRB\,171205A with the GRB/SN present. (c) Emission from the cocoon in the long GRB 171205A visible as an additional blue component at t$\sim$1\,day  \citep[from][]{Izzo19}.}
\label{fig:171205Ahost}
\end{figure}

At higher redshifts, the UV restframe gets shifted into the visible band, allowing to probe the restframe UV SFR. However, for low redshift GRBs ($z<1$), UV observations are important to directly probe the SFR and the UV luminosity function of GRB hosts. To date, this has only been done by extrapolating SED fits to galaxy templates obtained by observations in optical to NIR bands \citep[see also Fig. \ref{fig:171205Ahost} or][]{Schulze15}. 

An important goal for a UV space observatory, such as \emph{QUVIK}, is to build up a sample of low-redshift GRB hosts, something currently lacking in the field. Up to $z\sim0.5$ this is feasible for a significant part of the hosts, as can be seen from the \emph{QUVIK} limits indicated in Fig. \ref{fig:hostcontamination}. For fainter galaxies, deeper observations, reaching 1--2 mag fainter, would be beneficial. 
For the brighter galaxies and/or very nearby GRBs, late observations of the host will be needed in any case to subtract the host galaxy background (see Sec. \ref{subsectCONT}). 

\subsubsection{Cocoon emission of long GRBs}
The ejection of material from the interior of the star, swept up by the GRB jet, is responsible for the so-called cocoon emission \citep{Nakar17}. This emission is produced when the shock breaks out from the star, interacts with the surrounding medium and deposits energy. The exact strength depends on the mixing between the shock and the external medium.

This phenomenon was detected for the first time in GRB 171205A \citep[][see also Fig. \ref{fig:171205Ahost}]{Izzo19}, which was a low luminosity burst with a relatively weak afterglow. This GRB occurred at an unusually small distance of only 163\,Mpc, which was crucial to observe this feature. The cocoon was observed less than a day after the GRB as the additional black-body emission rapidly cooled and faded. Spectroscopic observations revealed material at very high speeds (1/3 of the speed of light) and rich in Fe and Ni, indicating it had to come from the very inside of the star. 

The cocoon emission is expected to be about two magnitudes brighter in UV compared to optical wavelengths \citep{Nakar17}.  It can also be observed at larger angles from the axis of the GRB jet, which would allow large FoV missions such as \emph{ULTRASAT} \citep{Sagiv14} to detect such signals even in the absence of a GRB\@. \emph{QUVIK} will be able to follow up \emph{ULTRASAT} detections in two complementary bands with a narrower PSF and look for re-brightenings in the UV within a day after a GRB, which could be indicative of a cocoon emission.

\subsection{FUV observations of GRBs \label{sect:FUV}}

\begin{figure}[t!]
    \renewcommand{\familydefault}{\sfdefault}\normalfont
    \centering
    \includegraphics[width=0.4\textwidth]{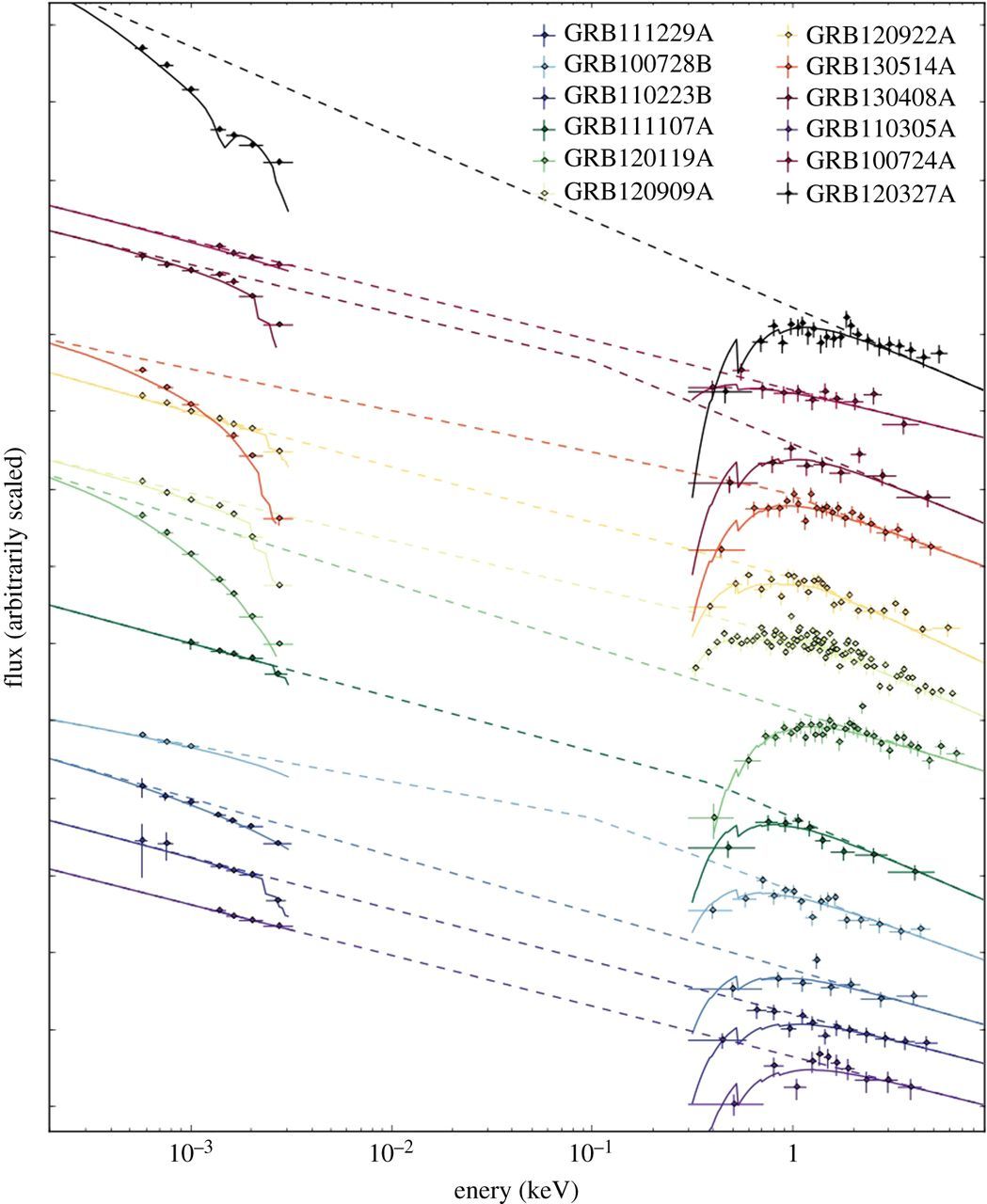}
    \includegraphics[width=0.59\textwidth]{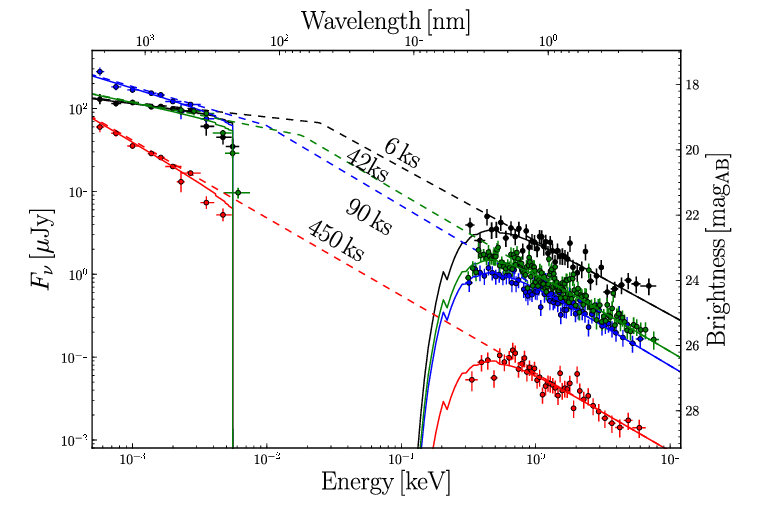}
    \caption{Left: Sample of GRBs with observations in the optical-UV and X-ray wavelengths, and fit to the spectral slopes for the two regimes \citep[from][]{SchadyREV}. Right: SED evolution of GRB 100418A from X-ray through UV and optical bands. The cooling break is shifting from the FUV towards the optical bands during the observations \citep[from][]{Nardini14}.}
    \label{fig:UVXray}
\end{figure}

To date, there have been no FUV observations of GRBs due to the lack of FUV observatories that allow for ToO observations; hence there is a gap of unknown SED between UV and X-rays (see Fig. \ref{fig:UVXray}). In the slow cooling regime, described in \cite{Sari98}, the cooling frequency can lie either in the X-ray regime or in the gap and is expected to evolve towards lower frequencies (longer wavelengths) as t$^{-0.5}$ \citep[see Fig. \ref{fig:sketch} and][]{Sari98}.  Due to the Lyman limit, the exact evolution of this gap is difficult to observe and follow, and its presence is inferred by a mismatch of the spectral slope between X-rays and UV-optical-NIR (see Fig. \ref{fig:UVXray}). However, these studies are limited to GRBs below $z\sim0.5$, above which they will start missing flux due to Lyman dropout. 

The FUV channel onboard \emph{QUVIK} will thus have an impact on several of the above-mentioned science areas: 

\begin{itemize}
    \item Improve the determination and evolution of the cooling break for low-redshift GRBs by extending the optical-UV SED further into the FUV.
    \item Observe early afterglow features of low-redshift GRBs such as flares or re-brightenings and better determine their behaviour in different bands; look for colour changes or a gradual shift towards shorter wavelengths. 
    \item Improve the photometric redshift estimates, which is especially crucial at low redshifts \citep[see e.g.][]{Kruehler11} and disentangle the effect of dust from the lack of flux due to the Lyman limit. 
    \item Observations of galaxy hosts in FUV do not exist so far but \emph{GALEX} observations of nearby field galaxies have proven to be valuable as indicators of unobscured star formation \citep{Morrissey2006}. The  FUV to NUV slope can also give an independent value for the extinction in case the stellar population of the galaxy is sufficiently known. 
    \item Allow for very early observations of the cocoon emission from the interaction of the jet and the surrounding medium. 
\end{itemize}

\subsection{Detectability of GRBs with \emph{QUVIK}}

\begin{figure}[h!]
    \renewcommand{\familydefault}{\sfdefault}\normalfont
    \centering
    \includegraphics[width=0.49\textwidth]{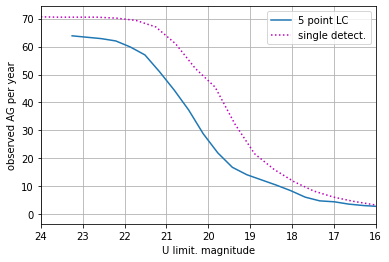}
    \includegraphics[width=0.49\textwidth]{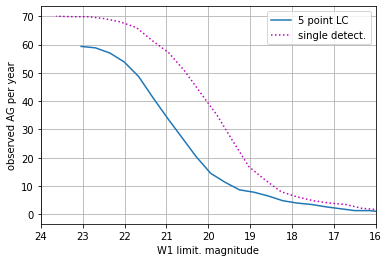}\\
    \includegraphics[width=0.49\textwidth]{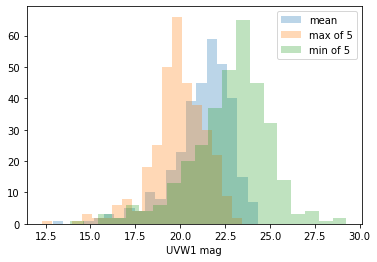}
    \caption{Top: Detection rate of GRBs in \emph{Swift}/UVOT \textit{u} and \textit{uvw1} bands.  Bottom: Brightness distribution for the average of the first five detections by UVOT split into the brightest, average, and faintest detections. All data are based on the second \emph{Swift} data release \citep{Roming2017}.}
    \label{fig:detection}
\end{figure}

We estimate the number of afterglows that could be observed by \emph{QUVIK} based on data from \emph{Swift}/UVOT published in the second UVOT catalogue \citep{Roming2017}.  This catalogue contains data and fitted parameters for 538 bursts over the first 6 years of \emph{Swift} observations. To estimate the sensitivity of \emph{QUVIK}, we use 2 bands based on UVOT filters:\footnote{\url{https://www.swift.ac.uk/analysis/uvot/filters.php}} \textit{u} filter centred at 346\,nm and \textit{uvw1} filter centred at 260\,nm.  Fig. \ref{fig:detection} shows the average number of bursts per year where either one observation or five data points reached a specific magnitude limit based on simple statistical calculations. About 6\% of the light curves in the catalogue have more than 40 points with better than 20\% precision in the \textit{uvw1} band.

Compared to \textit{Swift}/UVOT, \emph{QUVIK} will have the advantage of simultaneous observations in two bands.  
The on-target slew time will most likely be longer than for \emph{Swift}, however, the onboard GRB detector would reduce the reaction time significantly. For UVOT data in the \textit{uvw1} band, 35\% of light curves started $< 15$ minutes and 55\% in $< 1$ hour after the trigger.

Due to a typical power-law decay, the usual observation strategy of UVOT (similar to many ground-based follow-ups) is to increase the length of exposures as the afterglow becomes fainter. Grouping available data by exposure times from the two catalogues, we can derive the detection limits for the \textit{u}, \textit{uvw1} and \textit{uvw2} filters: half of the objects detected in a 500\,s exposure lies above 22.0, 21.8 and 22.3\,mag, respectively; changing the percentile to 90\% of detections (still 500\,s exposure), the limits change to 23.8, 23.6 and 24.4\,mag, respectively.  For a 20-minute exposure, the median detected magnitude is 22.6, 22.5 and 23.0, respectively; the faintest 10\% of the detections fall below 24\,mag for all filters.

Based on the \emph{Swift}/UVOT sample, our expected detection rate is $\sim$ 50 events per year with several data points on the light curve. For pure detection, we expect a rate of $\sim$ 70 per year, assuming a limiting magnitude of 22.

\subsubsection{Host galaxy contamination
\label{subsectCONT}}

\begin{figure}[h!]
    \renewcommand{\familydefault}{\sfdefault}\normalfont
    \centering
    \includegraphics[width=0.49\textwidth]{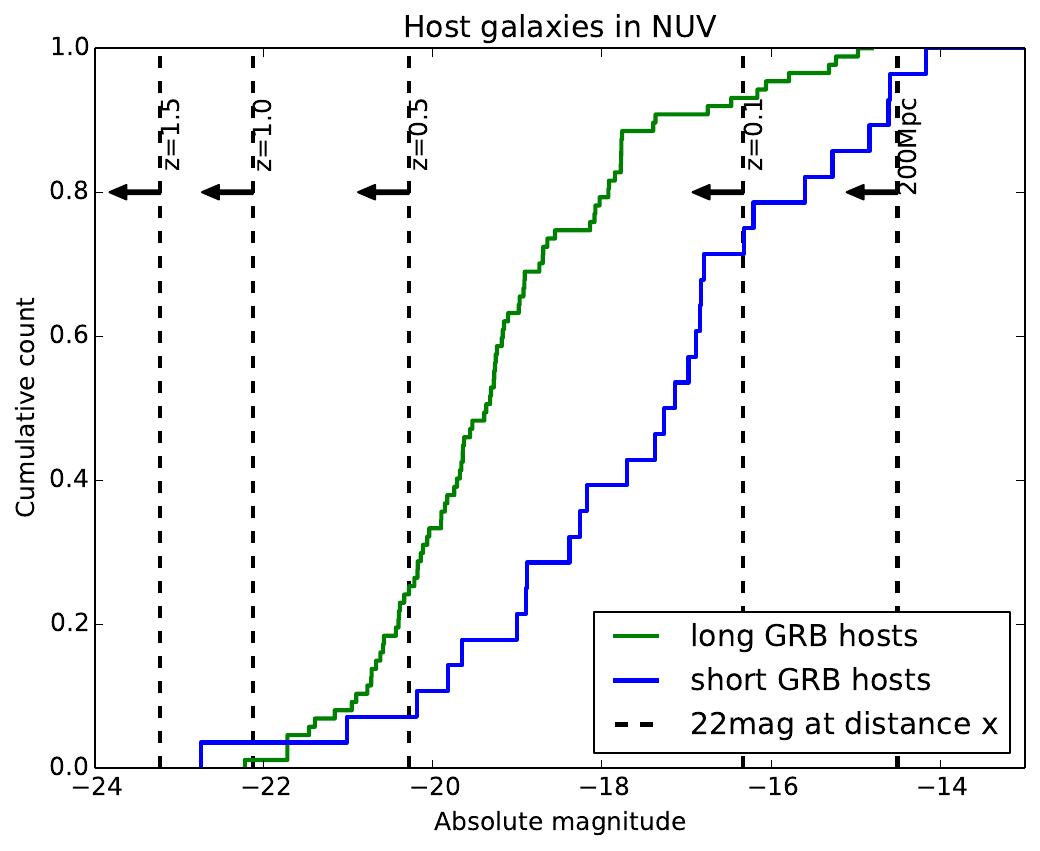}
    \includegraphics[width=0.49\textwidth]{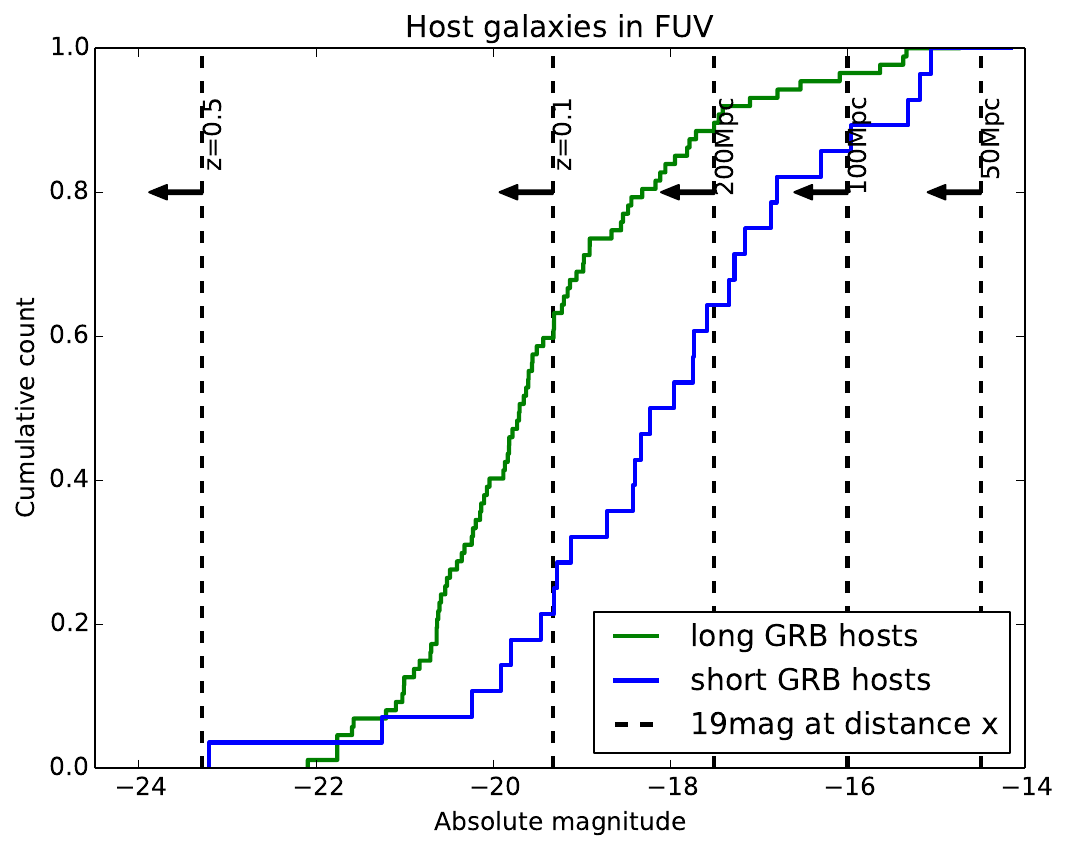}
    \caption{Global host galaxy absolute magnitudes extrapolated from SED fits based on optical to NIR data. Dashed lines show the absolute magnitude corresponding to an observed magnitude of 22 in NUV and 19 in the FUV band at different redshifts.}
    \label{fig:hostcontamination}
\end{figure}

Contamination by the host galaxy can be an important issue for GRB afterglows. Long GRBs have a very small offset from their host galaxy with an average offset of 0.6 $R_{\mathrm{e}}$ and 80\% being within $1 R_{\mathrm{e}}$ \citep{Lyman17}. The average effective radius is only $1.7\pm 0.2$\,kpc. Short GRBs have larger offsets of $\sim 1.5 R_{\mathrm{e}}$ \citep{fong2022} but also larger hosts with an average radius of 3.3\,kpc.

To estimate possible contamination, we used SED fitting of a sample of short and long GRB hosts with data in the optical and NIR and extrapolated the result to the NUV and FUV bands. The NUV and FUV filters are based on the passbands of the \emph{GALEX} UV satellite, where the FUV covers the 134--180\,nm range and the NUV spans the 170--300\,nm band. The data of the SED fits are based on the SHOALS sample for long GRB hosts \citep{Perley16} and the samples of \citet{fong2022} and \citet{Nugent22} for short GRBs (Ag\"u\'i Fern\'andez et al. in prep.).

The distributions for NUV and FUV are shown in Fig. \ref{fig:hostcontamination}.  As expected, long GRB hosts are on average brighter in the UV than short GRB hosts since they are actively forming stars and the average offset of a long GRB from its host ($0.6 R_\mathrm{e}$) is relatively small. 
 With the larger offset for short GRBs, contamination might be less of an issue even at lower redshifts; however, short GRB afterglows are also on average fainter \citep{Kann11} and hence more often close to our detection limit. These results show that for most low-redshift GRBs, we will need an additional image of the host galaxy, obtained at a later time, for image subtraction.

\section{The UV emission of fast radio bursts and magnetars}

Fast radio bursts (FRBs) are extraordinarily bright millisecond transients first 
observed in the radio band \citep{Lorimer2007,Zhang2020}.
Some FRBs were found to be repeating and their association with distant 
galaxies confirmed their extra-galactic origin. 
Importantly, their repeating nature indicates that 
their progenitors are non-cataclysmic, although it is not yet clear whether 
all FRBs are repeating and whether they all have the same progenitors \citep{Kramer2023}. 
To date, more than 1000 individual (``one-off'') FRBs events and about 50 
repeating sources have been detected, but their exact origin remains a mystery \citep{Zhang2020}. 
Even though it is not clear if all FRB sources come from a unique population of objects,
the mainstream set of models link FRBs with magnetar activity.

\begin{figure}[!ht]
    \centering
    \includegraphics[width=0.5\textwidth]{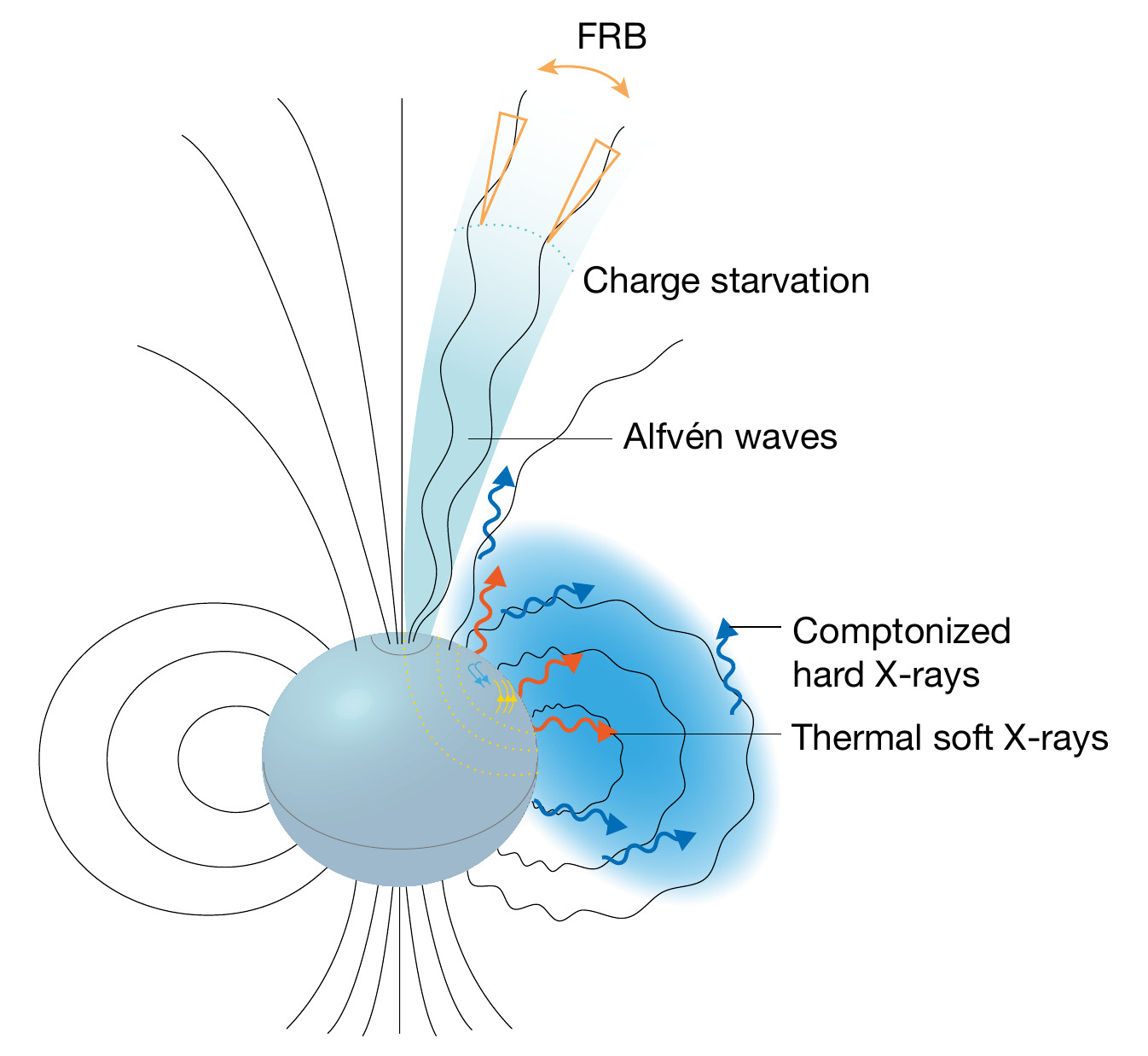}
    \caption{Radiation model of FRBs that includes the magnetosphere of the neutron star \citep{Zhang2020}.}
    \label{fig:FRBmodel}
\end{figure}

Recently, the soft gamma-ray repeater SGR~1935+2154 was observed to be associated with both an FRB \citep[although its radio luminosity is orders of magnitude smaller than that of extragalactic FRBs;][]{CHIME2020,Bochenek2020} and a simultaneous X-ray flare \citep{Ridnaia2021,Tavani2021,Mereghetti2020,Li2021}, becoming the first FRB detected in the Milky Way.
Soft gamma-ray repeaters are magnetars, neutron stars with extremely strong magnetic fields (10$^{13}$--10$^{15}$\,G), showing X-ray and soft gamma-ray bursts.
This observation suggests that they emit the radio and X-ray emissions simultaneously \citep{CHIME2022}.
The FRB emission model of a magnetar is shown in Fig.~\ref{fig:FRBmodel}.
Nonetheless, there are also many other models for FRB sources, for example, coherent radio bursts of relativistic shocks \citep{Platts2019,Zhang2020}.

Many theoretical models have been proposed to interpret FRBs in various spectral bands \citep{Platts2019}; however, the exact emission mechanism is so far unknown.
UV measurements can open a unique spectral window to place constraints on the emission properties of magnetars producing FRBs and to refine theories of their emission mechanisms.
Assuming that FRBs are indeed produced by magnetars, an observation performed during or just after a soft gamma-ray/X-ray flare could, in principle, place constraints on their physics.
Such a measurement of a UV flux of an FRB counterpart has a major breakthrough potential because no simultaneous X-ray--UV or radio--UV detection has been obtained so far.

\begin{figure}[!ht]
    \centering
    \includegraphics[width=0.6\textwidth]{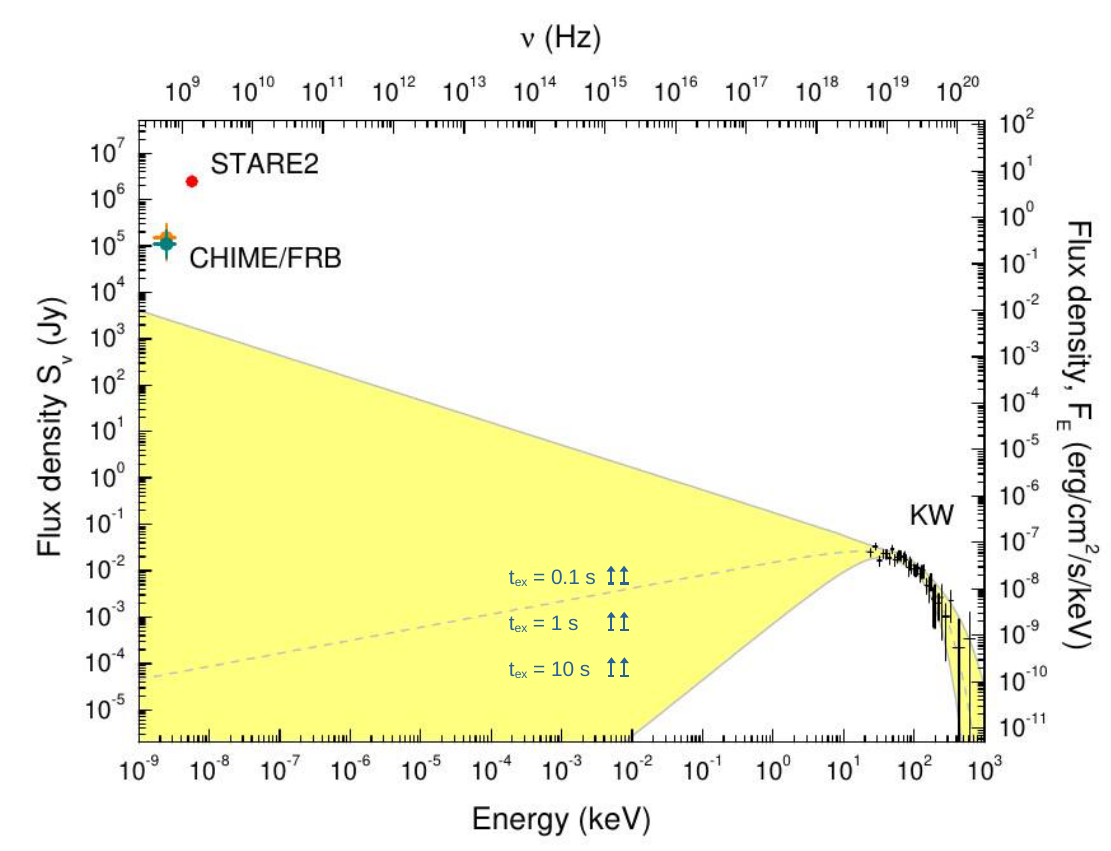}
    \caption{Average spectral distribution of SGR~1935+2154, the proposed source of FRB~200428  
    observed by the Konus-\textit{Wind} (KW) instrument. The detected FRB fluxes are shown 
    in observations of CHIME/FRB (green, orange) and STARE2 (red) \citep{Bochenek2020}. 
    The yellow region denotes the probable detection interval, assuming the fitting 
    model of the X-ray spectrum. The blue arrows indicate the 5$\sigma$ detection threshold of a single exposure for exposure times 0.1\,s, 1\,s, and 10\,s.
    From \citet[][]{Ridnaia2021}.}
    \label{fig:NS1}
\end{figure}

A potential candidate for observation by \emph{QUVIK} is the Galactic magnetar 
SGR~1935+2154, which has previously been associated with the above-mentioned 
low luminosity FRB (Fig.~\ref{fig:NS1}, \citealt{Ridnaia2021}). This soft 
gamma-ray repeater in a flaring state will present a perfect opportunity for 
long-term monitoring.
Even though an FRB is a millisecond event in radio, magnetar bursts, outbursts, and giant flares  
can last milliseconds to days \citep{Kaspi2017} in X-rays and can be followed by an afterglow. 
Thus, the emission model also predicts a longer-lasting increase in the UV emission. 
In addition, there is a known delay in X-ray and radio observations of several 
milliseconds for the Galactic magnetar SGR~1935+2154 and the associated burst 
FRB~200428 \citep{Ge2023}. 
As can be estimated from the spectral fit in Fig.~\ref{fig:NS1},
observations could reach intensities $\approx 5000\,\mu$Jy,
$\approx 500\,\mu$Jy, and $\approx 50\,\mu$Jy for 0.1\,s, 1\,s, and 10\,s
exposure times, providing important constraints on the emission model. 
In addition, the magnetar activity time window can be much longer than the assumed exposure times.
For the long-time activity windows, sources brighter than $\sim$23.5\,mag 
can be detected in up to several hours long exposure time.
Nonetheless, in this case, also other UV satellites such as {\it HST} or {\it AstroSat} could be used.

Simultaneous UV and radio observations of nearby repeating FRB sources could provide valuable measurements or upper limits on their UV fluxes.
The repeating sources could be the first targets of FRB observations because of their high repetition level. The nearest known repeating FRB source in a globular cluster of the M81 galaxy at a distance of 3.6\,Mpc demonstrated remarkable activity when it emitted 53 bursts within only 40\,min \citep{nimmo2023}.

A UV detection of a magnetar before, during, or after an FRB will help to distinguish 
between different emission models. Combining the UV data with those 
in X-rays allows for spectral fitting of various emission models because a wider 
frequency range is covered. For example, \citet{Ridnaia2021} used a cutoff power 
law and a sum of two blackbody functions for fitting the X-ray spectrum.
However, more complex profiles may appear when the UV band is 
included. Such broadband coverage could reveal whether the radiation is 
produced by a thermal or nonthermal emission mechanism or their combination.
While blackbody radiation can be expected to be connected to the neutron 
star surface \citep{Riley2019}, nonthermal spectra can originate from 
magnetospheric plasma \citep{Cerutti2016}. The study of the emission of magnetars 
in their quiet state is described in Paper II \citep{quvik2ssrv}.

\section{UV emission of supernovae}%
\label{sec:supernovae_novae}

The physics of explosions of massive stars as supernovae (SNe) remains one of the outstanding unsolved problems in astrophysics. Early UV observations of SNe present a very effective tool for studying the radius, surface composition, as well as the surrounding circum-stellar medium (CSM) of massive stars that eventually explode as SNe \citep[e.g.][]{2017suex.book.....B}, providing key initial conditions for computer simulations of the explosive process. Here, we focus on observations of the initial 
shock breakout, as well as on type IIn and super-luminous SNe (SLSNe), which may provide some key ingredients to SN study.

\subsection{Supernova shock breakout}

Observations of the so-called shock breakout, which is usually regarded as the start time of an SN, require measurements in X-ray and UV wavelengths obtained in the first hours and up to a day after the explosion and could be triggered by {\emph {ULTRASAT}} or by ground-based wide-field surveys. The SN shock breakout emission may provide invaluable information about the progenitor size and properties of the surrounding CSM, including its asymmetry, inhomogeneities, asphericity \citep[e.g.][]{2022ApJ...931...15B}, and chemical composition 
\citep[e.g.][]{2006ApJ...644L.171N, 2017ApJ...840...57Y}.

The shock breakout is defined as the moment when the internal SN shock wave emerges from the progenitor's photosphere \citep[e.g.][]{2010ApJ...725..904N, 2017hsn..book..967W}. After escaping the stellar body and entering the CSM surrounding the progenitor star, the shock wave accelerates and produces X-rays accompanied by a longer-lasting afterglow in the UV. 
As demonstrated by e.g.~\citet{2008ApJ...683L.135C} and \citet{2008Natur.453..469S} for SN 2008D, X-ray and UV observations of the early characteristics can be used to determine the radius of
the SN progenitor. An equally important goal is to map the morphology and properties of the transition zone between the star and the surrounding CSM. The density and velocity profiles of this region remain not well-understood. The same applies to the time variability of these quantities, even if we assume only spherically symmetric explosions.  We expect that UV photometry will contribute significantly to our understanding of the non-stationary, eruptive pre-explosive ejections and enhanced mass loss.

\begin{figure}[!t]
    \centering
   \includegraphics[width=7cm]{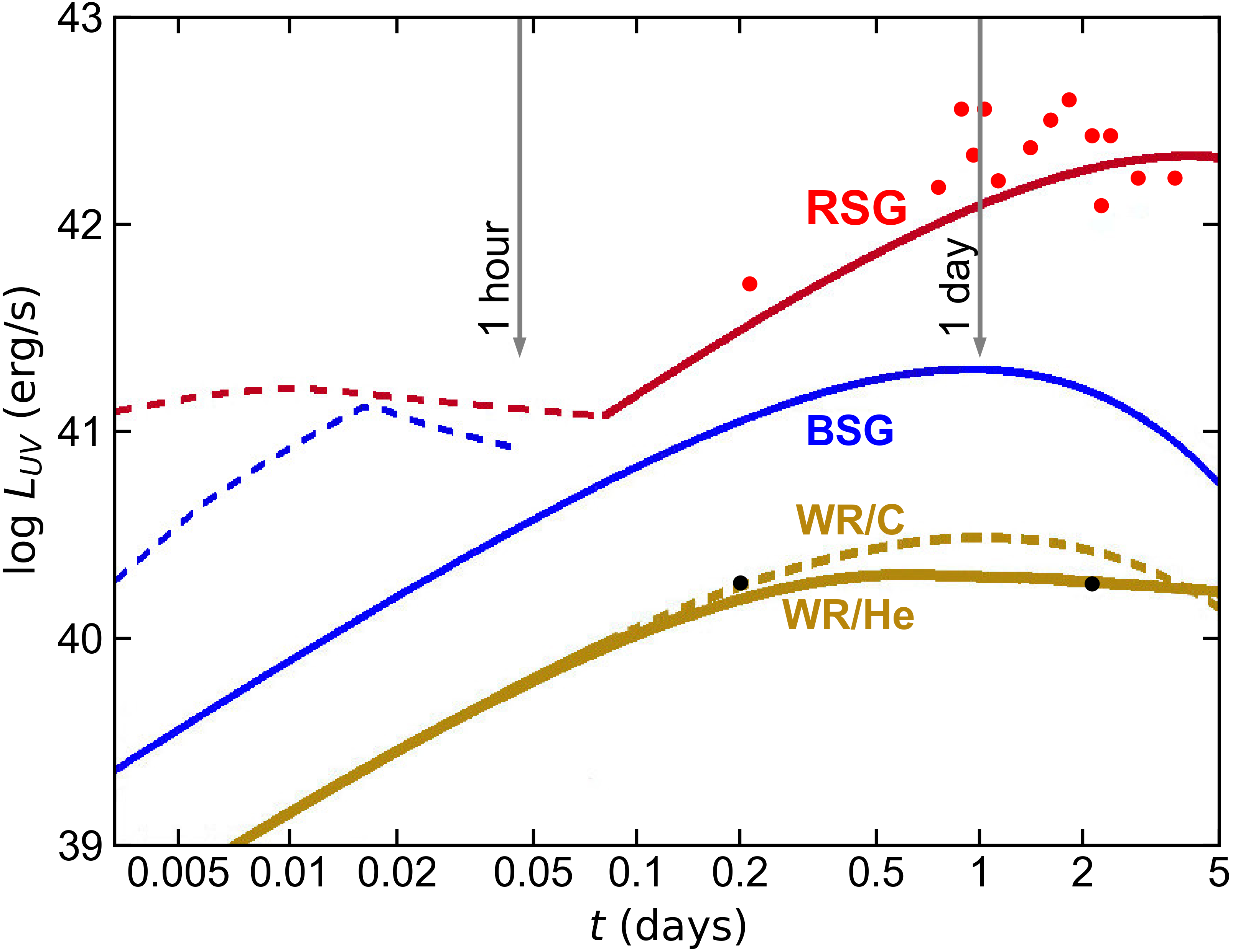}
    \caption{Illustration of early UV emission models of SN compared with observations. Models of \citet{2013ApJ...774...79S}---red supergiant (RSG; dashed red line) and blue supergiant (BSG; solid blue line); \citet{2010ApJ...725..904N}---BSG, dashed blue line; \citet{2011ApJ...728...63R}---RSG, solid red line, Wolf–Rayet (WR) star with He, solid yellow line, and C/O dominated WR star, dashed yellow line.  Observations from GALEX/NUV \citep['supernova legacy survey' --- SNLS-04D2dc, type II-P, ][red circles]{2008Sci...321..223S, 2008ApJ...683L.131G} and Swift/UVOT \citep[SN 2008D, type Ib,][black circles]{2008Natur.453..469S}. Inspired/adapted from \citet{2016ApJ...820...57G}.\label{fig:ugal}}
\end{figure}

The current lack of X-ray/UV measurements does not allow us to determine the characteristic chemical composition and plasma parameters of SN explosions with the required accuracy. To a large extent, it prevents the proper classification of SNe and the study of the origin of elements \citep{2017hsn..book..967W}. Namely, in FUV, thousands of spectral lines of iron group elements are the source of significant absorption. Therefore, the FUV excess may help us to reveal the abundance of the freshly synthesised iron-peak group elements in the outer ejecta and relate it to the overall metal abundance of the SN progenitor \citep{2017ApJ...840...57Y}. 
 
However, the approach described so far does not include the geometric and physical complexity imprinted in this early-phase emission \citep[e.g.][]{2010ApJ...725..904N}.  Namely, it is becoming increasingly evident that the asphericity of the expanding SNe envelope, the surrounding CSM, and the star-CSM transition region play a crucial role in the investigation of the pre-SN and SN phenomena \citep[e.g.][]{2011ApJ...727..104C, 2017hsn..book.1791C, 2017ApJ...837...84J}. Asymmetries may induce a great number of secondary shocks of various strengths and directions and thus significantly alter the timescale of the shock breakout and the related high-energy afterglows \citep[e.g.][]{2013ApJ...779...60M, 2020ApJ...898..123F, 2021MNRAS.508.5766I}. Moreover, the 
multiplicity in time of such shocks can dramatically change the overall observational picture of SNe. The shocks may thus help to model and understand the conditions and processes in the progenitor star that may have led to the overall mass, temporal variability,
and spatial complexity of the CSM ejection. The complexity of the measurement is, of course, enhanced by the fact that the emission resulting from the SN-CSM interaction takes place in many wavebands. Thus, a thorough examination and comparison of observations performed in different wavelengths will help us to decipher the processes behind this emission.

For these reasons, the exact behaviour of UV light curves is somewhat heterogeneous; their detailed time evolution and luminosity depend both on the above-mentioned characteristics and on the particular SN type. To determine the common features of the UV luminosity behaviour, we can take, for example, the models of \citet{2011ApJ...728...63R}, \citet{2010ApJ...725..904N}, and others, together with observations of, e.g. \citet{2008Sci...321..223S}, \citet{2008ApJ...683L.131G}, \citet{2008Natur.453..469S}, shown in Fig.~\ref{fig:ugal}. FUV light curves show an initial rise over a duration
$R_\star/c$ (where $R_\star$ is the radius of a progenitor and $c$ is the speed of light), which corresponds to about 2000\,s after the shock breakout for a red supergiant (RSG) explosion and correspondingly less for a blue supergiant (BSG) progenitor. However, this fast rise is followed by a slow decay up to typically 14\,hrs for an RSG and 0.5--1\,hr for a BSG (assuming that the breakouts from RSGs and most of BSGs are in thermal equilibrium). This decrease is followed by a second peak, as the "observed temperature" falls into the given frequency range, typically at 2\,days for RSGs, and slightly less than 1\,day for BSGs, while it occurs at a yet earlier time for more compact progenitors \citep[see, e.g. Fig.~4 in][]{2010ApJ...725..904N}. The peaks and profiles of NUV light curves evolve more slowly, on the order of days after a shock breakout (see Fig.~\ref{fig:ugal}). Considering \emph{QUVIK}'s expected slew time of about 20 minutes, observing these light curves after receiving an alert from a wide-field transient survey \citep[for example, the anticipated alert time of \emph{ULTRASAT} is 15 minutes;][]{Shvartzvald2023} appears feasible.  

\subsection{Early-to-intermediate time emission of super-luminous supernovae}

\begin{figure}[!t]
    \centering
    \includegraphics[width=6.75cm]{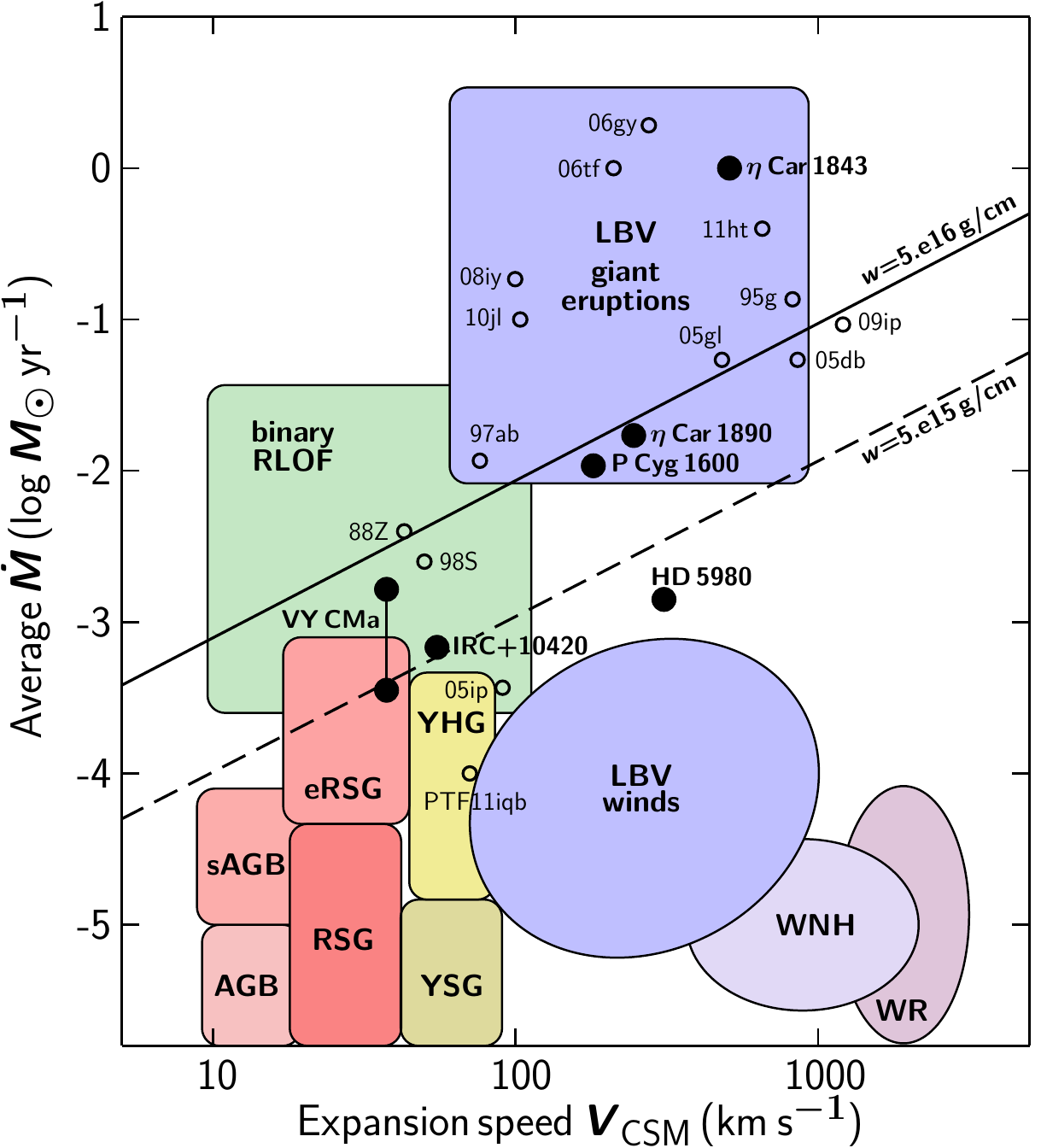}
    \caption{Schematic plot of average mass-loss rate of relevant types of stars with high mass-loss rates versus the velocity of their winds or eruptive mass-losing events. The coloured
    regions highlight the values of selected types of stars or their configurations, that is (from bottom left) of asymptotic giant branch (AGB) and super-AGBs (sAGB), red supergiant (RSG) and extreme RSG (eRSG) stars, yellow supergiants (YSG), yellow hypergiants (YHG), luminous blue variable (LBV) winds, WN stars with H (WNH), H-free WR stars, binary Roche-lobe overflow (RLOF), and LBV giant eruptions.  Several famous stars with extremely high mass-loss rates are plotted with larger black solid circles, while the smaller empty circles localize the other well-known examples of SNe IIn.  The two diagonal lines increasing to the upper right are the wind density parameters ($w=\dot M/V_\text{CSM}$), denoting the typical (solid line) and the lower (dashed line) values required for producing SNe IIn.  Some specific stars may `sit' below these limits in case of significantly aspherical or clumpy winds, etc. Inspired by \citet{2017hsn..book..403S}.}%
    \label{fig:Smith1}
\end{figure}

Important targets for \emph{QUVIK} are type IIn SNe \citep{1990MNRAS.244..269S, 1997ARA&A..35..309F, 2014ARA&A..52..487S, 2017hsn..book..195G}, which show particularly strong and narrow-to-intermediate Balmer emission lines. The narrow lines, usually superimposed on much broader lines that manifest as wings, are formed in the photoionized, relatively slow unshocked material surrounding the SN before the explosion; they reveal SNe that expand into a dense pre-explosion CSM \citep[e.g.][]{1982ApJ...258..790C, 2013MNRAS.428.1020M, 2016MNRAS.458.2094D, 2020A&A...642A.214K}. Fig.~\ref{fig:Smith1} illustrates the required mass-loss rates of stellar types that are considered to supply the CSM by a relevant amount of steady or eruptive mass ejection before the SN event, demonstrating the importance of the pre-explosion CSM asphericity.

It is still difficult to judge whether SN IIn will manifest as 
super-luminous supernovae (SLSNe).
 The class of SLSN defined by \citet{2012Sci...337..927G} exhibits extremely bright and long-lasting light curves with peak absolute magnitudes of $-21$\,mag or above in the visible band.  Their duration may be hundreds of days or even more than a year. This may also be accompanied by various bumps or rebrightenings in the bolometric light curve as well as in particular photometric bands \citep[e.g. Fig. 10 in\,][]{2020A&A...642A.214K}. 
SLSNe are divided into types I and II according to the presence of hydrogen in the spectrum. Light curves of SLSNe I usually evolve relatively quickly, with a rise time of several weeks to more than a month, which is somewhat shorter than in SLSN II, followed by a relatively fast decline. SLSN I are of significant interest because of their association with LGRBs and KNe, as well as their large UV luminosities \citep[][see also Fig.~\ref{fig:brownie}]{2017suex.book.....B}.
Their extreme luminosities and long durations
(in particular for SLSNe II)
 are mostly interpreted as a result of interaction with extended and inhomogeneous CSM. Still, other engines, such as magnetars, might also play a role \citep{2017ApJ...841...14M}. According to the extent, total mass, morphology, density distribution, or chemical composition of the CSM surrounding the explosion, the characteristic narrow Balmer line occurrence may last for days, weeks, or even years \citep{2020ApJ...899...51S}.  Well-documented examples of the long-term objects of this type may be the prominent SNe: SN1988Z \citep{2002ApJ...581..396W, 2006ApJ...646..378S, 2017MNRAS.466.3021S}, SN1998S \citep{2000MNRAS.318.1093F, 2001MNRAS.325..907F}, SN2005gl \citep{2007ApJ...656..372G}, SN2010mc \citep{2013Natur.494...65O}, or SN2010jl \citep{2014Natur.511..326G, 2014ApJ...781...42O}.

Magnetar-driven shocks, which may be important in SLSNe, may differ from the shock breakout of a 'normal' SN accompanied by a brief X-ray burst \citep[e.g.][]{1978ApJ...223L.109K, 1999ApJ...510..379M,2016ApJ...821...36K}. Since the magnetar-inflated nebula strongly accelerates the surrounding SN ejecta \citep{2017ApJ...841...14M}, a 
weakened magnetar-driven shock will radiate at longer wavelengths (FUV/NUV/optical), and its emission will last longer. The double-peaked early light curve will also indicate the presence of a magnetar-driven SN due to the persisting injection of energy into the shell. That will produce an unusually fast shock that can radiate outside the bulk of the SN ejecta before most of the centrally thermalised energy has had time to diffuse out \citep{2010ApJ...717..245K, 2016ApJ...821...36K}.

An emerging class of SN-like explosions with luminosities $\gtrsim10^{44}$\,erg\,s$^{-1}$, which may even exceed SLNSe, are the "Fast Blue Optical Transients" \citep[FBOTs;] []{drout2014,prentice2018,ho2020,metzger2022}. The prototypical member of this class is AT2018cow, the optical emission of which rose over a few days to a luminosity of $L\approx4\times10^{44}$\,erg\,s$^{-1}$. 
The physical nature of this phenomenon, especially in the case of its exceptionally "luminous" subclass (LFBOTs), could be associated with the tidal disruption and hyper-accretion of a massive hydrogen-depleted Wolf–Rayet (WR) star by a black hole or neutron star binary companion. The optical/UV flux tends to be stable for these types of transients; for example, for AT2018cow, it showed a change of only about 0.1\,mag over 1000 days \citep{metzger2022}. 
The study of this new class of fast, UV-bright transients is expected to benefit strongly from the quick 2-band photometry provided by \emph{QUVIK}. 

To summarise, potential science objectives of a small two-band NUV/FUV space telescope observing the early-to-intermediate time SN emissions include: 
\begin{itemize}
    \item[\textbullet]The shocks that propagate in the CSM after the shock breakout \citep{2010ApJ...724.1396O} are predicted to radiate mostly in the UV and X-rays \citep{2011arXiv1106.1898K, 2011PhRvD..84d3003M, 2014MNRAS.440.2528M, 2012ApJ...747L..17C, 2020ApJ...899...51S}. Observation of the explosion and subsequent SN-CSM interaction at these wavelengths may reveal valuable information regarding the explosion mechanism, its morphology and chemical composition, and CSM global properties \citep[e.g.][]{2013ApJ...763...42O}. Measurement of the luminosity in the UV band will contribute to a much more precise estimate of the overall bolometric luminosity.\\
    \item[\textbullet]UV observations will help to determine the CSM geometry. Studies of well-known potential SN IIn progenitors \citep[e.g.~$\eta$ Car,][]{1997ARA&A..35....1D, 2012Natur.486E...1D, 2009Natur.458..865G} indicate that SN and CSM asymmetries must be taken into account to achieve realistic models of observed characteristics in the case of such physically and geometrically complex processes.\\ 
   \item[\textbullet]FUV measurements will provide deep insight into the very early light curves that may indicate weaker shocks due to the presence of magnetars or reveal a rapidly dimming phase of the radiative tail closely following the X-ray breakout flash \citep{2016ApJ...821...36K}. Furthermore, FUV observations may also help to reveal the chemical composition of SN ejecta \citep{2017ApJ...840...57Y}.
\end{itemize}

A realistic objective of a small, two-band NUV/FUV space observatory is to provide UV photometry of SNe in host galaxies within $z\le 0.1$. A real-time alert can be provided by soft X-ray and UV instruments with very large fields of view, such as the MAXI X-ray monitor on the ISS\footnote{\url{https://iss.jaxa.jp/en/kiboexp/ef/maxi/}} and \emph{ULTRASAT}, as well as by ground-based wide-field optical transient surveys and neutrino observatories\footnote{ \url{https://gcn.nasa.gov/notices}}.

\subsection{Late phase of Type II super-luminous supernovae}

\begin{figure}[!t]
    \centering
    \includegraphics[width=\textwidth]{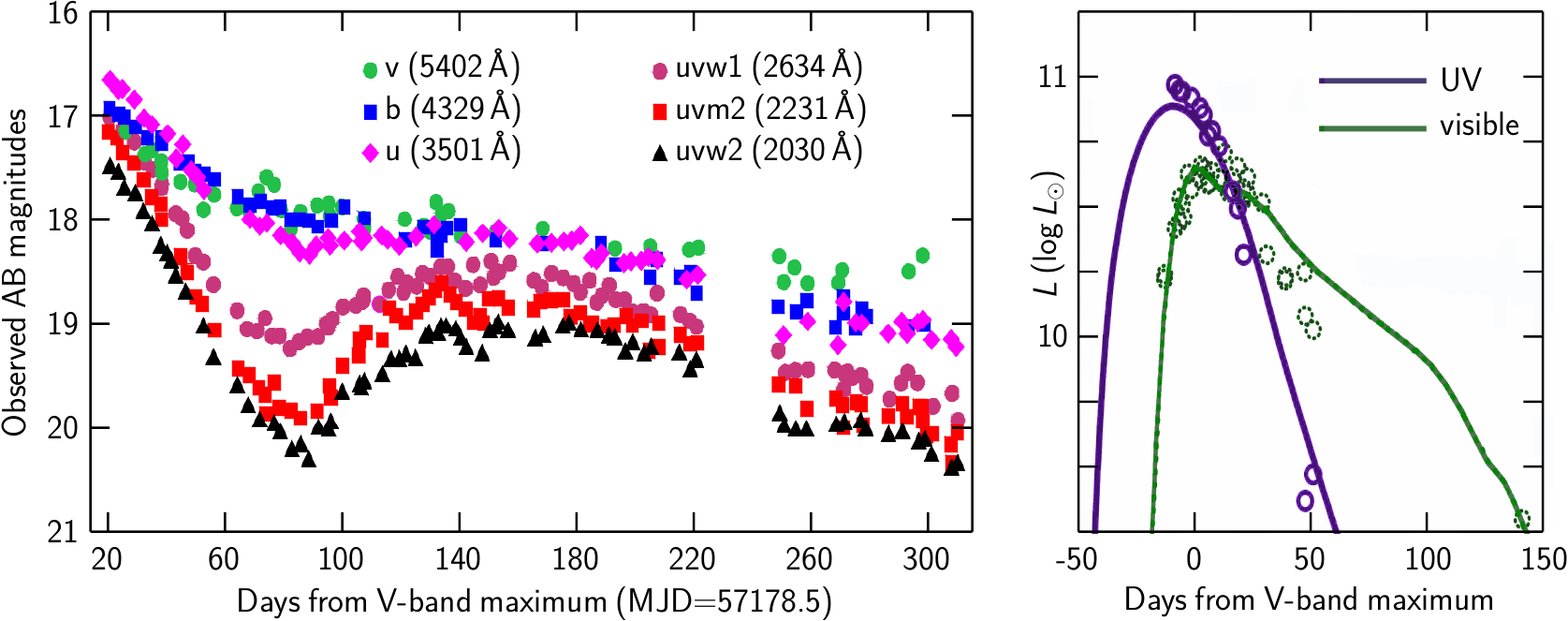}
    \caption{Left panel: Example of late-time SN emission: UVOT light curves of ASASSN-15lh 
    (major features characterise it as SLSN I) 
    in AB magnitudes.  The late time rebrightening is brighter than the $M < -21$\,mag cut-off for SLSNe.  The readjusted figure is adapted from \citet{2016ApJ...828....3B}. Right panel: UV and V-band light curves of SLSN I Gaia16apd (circles) plotted together with synthetic light curves from models of SN-CSM interactions (solid lines), reconstructed from \citet{2017ApJ...845L...2T}.}%
    \label{fig:brownie}
\end{figure}

An essential characteristic of SLSNe is their high UV luminosity (see Fig.~\ref{fig:brownie}). Therefore, their UV observations are crucial for determining their total luminosities and temperatures. UV observations of 
SLSNe II 
(and SNe IIn, which usually precede them and are related to them)
at later times are important for probing the complex CSM distribution in the wider surroundings of exploding massive stars.  Figure 10 in \citet{2020A&A...642A.214K} compares the models of asymmetric SN-CSM interactions with the observed bolometric light curves of selected SNe \citep{2014A&A...569A..23K, KKapplmath,2018A&A...613A..75K}.
 
Most of the SNe are of type IIn and show a steep initial decline followed by a long slow decline. Undulations and bumps in their light curves are rare but have been observed in a few cases 
\citep{2017A&A...605A...6N},
and their exact nature is still debated. The interaction of SN ejecta with a clumpy CSM \citep{2016MNRAS.455.4388C, 2020MNRAS.493..447C} is also expected to produce bumps in the light curves.

\section{Summary}

Recent progress in detector and coating technologies, the advent of high-quality off-the-shelf micro/small-satellite platform sub-systems, and the availability of affordable launch opportunities enable, for the first time in history, also smaller countries and organisations to deploy sensitive and highly responsive space observatories. The \emph{QUVIK} mission, described in Sect. \ref{sec:mission}, will enable significant progress in the study of explosive transients, as well as in the field of stellar astronomy \citep[see companion paper II,][]{quvik2ssrv} and in the study of active galactic nuclei \citep[see companion paper III,][]{quvik3ssrv}. 

The mission's primary objective is the follow-up of BNS and NS-BH mergers detected by gravitational wave observatories and wide-field surveys and their photometry in the UV band to discriminate between different kilonova scenarios. Their photometry obtained less than $\sim$6 hours after the merger will provide important constraints on the beta-decay of free neutrons. Early two-band photometry will also provide key data on shock interactions. The presence of early emission from beta-decay and shock interactions may increase the early UV brightness of kilonovae by as much as 2\,magnitudes, potentially significantly boosting their detectable number. The measurements performed by \emph{QUVIK} will also provide constraints on the structure and composition of the ejecta, thus probing the contribution of BNS and BH-NS to r-process nucleosynthesis. 

Simultaneous NUV/FUV photometry of GRB afterglows will enable to study and determine the nature of flares, rebrightenings and plateaus at the onset of the afterglow emission. The two UV bands will help to better understand the jet physics and the emission mechanism by searching for colour changes.
Two-band UV photometry of the rare cocoon emission feature from the interaction between the jet and the material of the star predicted by \cite{Nakar17} and observed by \cite{Izzo19} will be particularly valuable.  

Kilonova and GRB science will greatly benefit from a GRB detector with a fast onboard localisation capability. Such a detector will enable much-needed early follow-up of GRB afterglows and, in the case of some long GRBs, the detection of the prompt emission in two simultaneous UV bands. 

A sample of UV observations of kilonova and low-redshift GRB host galaxies, which is currently lacking, will be valuable for determining the ancestors of the progenitors.  

For supernovae, early observations of the shock breakout will provide information about the progenitor stars and their surrounding CSM. For super-luminous supernovae, later observations will also probe the CSM and the wider surroundings of massive exploding stars. Wide-field transient surveys operating or coming online in various wavebands are also expected to discover new classes of supernovae and other transients. The two-band UV-photometry mission thus provides opportunities for discoveries and follow-up observations of new types of yet unknown objects and phenomena. 

The upcoming democratisation of space activities is expected to result in a growing number of increasingly capable small satellites complementing large missions. The relative affordability and potential economies of scale may even result in greater numbers of small space observatories covering complementary parts of the electromagnetic spectrum or dedicated to particular targets or classes of objects. They are expected to be especially valuable for transient astronomy, but other fields will also benefit from the progress. \emph{QUVIK} may thus help pave the way for a future multi-wavelength constellation of space observatories monitoring the energetic universe.
\paragraph{Acknowledgement}
We thank the Czech Ministry of Transportation and the European Space Agency for their support of the \emph{QUVIK} project within the Czech Ambitious Mission Programme. J.B. acknowledges the support of the German Science Foundation (DFG) project BE~7886/2-1. G.L.-G. has been supported by the fellowship Lumina Quaeruntur No. LQ100032102 of the Czech Academy of Sciences. The research of OP has been supported by Horizon 2020 ERC Starting Grant `Cat-In-hAT' (grant agreement no. 803158). We thank B. D. Metzger for his valuable comments.

\section*{Declarations}

\paragraph{Competing Interests} The authors declare no competing interests.

\bibliography{sn-bibliography}

\bibliographystyle{aa} 

\end{document}